\documentclass[journal]{IEEEtran}
\ifCLASSINFOpdf
\usepackage[pdftex]{graphicx}
\else
\fi
\usepackage{bm}

\usepackage{multicol}
\usepackage{lipsum}
\usepackage{amsmath}
\usepackage{cite}
\begin{document}
%
\title{Spectral ghost imaging camera with\\ super-Rayleigh modulator}
%
%
%

\author{Shengying~Liu,
        Zhentao~Liu,
        Chenyu~Hu,
        Enrong~Li, 
        Xia~Shen
        and~Shensheng~Han
\thanks{This work was supported by Youth Innovation Promotion Association of the Chinese Academy of Sciences, and Defense Industrial Technology Development Program of China (D040301), and National Natural Science Foundation of China (61571427). (Corresponding author: Zhentao Liu)}
\thanks{Shengying Liu, Zhentao Liu, Chenyu Hu, Enrong Li, Xia Shen and Shensheng Han were with  the Key Laboratory for Quantum Optics and Center for Cold Atom Physics, Shanghai Institute of Optics and Fine Mechanics, Chinese Academy of Sciences, Shanghai 201800, China and Center of Materials Science and Optoelectronics Engineering, University of Chinese Academy of Sciences, Beijing 100049, China (e-mail: shengyl@siom.ac.cn;ztliu@siom.ac.cn; huchenyu@siom.ac.cn; ler@siom.ac.cn; shenxia@siom.ac.cn; sshan@mail.shcnc.ac.cn).}
}

%
%

\markboth{Journal of \LaTeX\ Class Files,~Vol.~14, No.~8, August~2015}%
{Shell \MakeLowercase{\textit{et al.}}: Bare Demo of IEEEtran.cls for IEEE Journals}
%



\maketitle

\begin{abstract}
A spectral camera based on ghost imaging via sparsity constraints (GISC) acquires a spectral data-cube $(x,y,\lambda)$ through a single exposure. The noise immunity of the system is one of the important factors affecting the quality of the reconstructed images, especially at low sampling rates. Tailoring   the intensity to generate super-Rayleigh speckle patterns which have superior noise immunity may offer an effective route to promote the imaging quality of GISC spectral camera. According to the structure of GISC spectral camera, we proposed a universal method for generating super-Rayleigh speckle patterns with customized intensity statistics based on the principle of reversibility of light. Simulation and experimental results demonstrate that, within a wide imaging spectral bandwidth, GISC spectral camera with super-Rayleigh modulator not only has superior noise immunity, but also has higher imaging quality at low sampling rates. This work will promote the application of GISC spectral camera by improving the quality of imaging results, especially in weak-light illumination.
\end{abstract}


%
\IEEEpeerreviewmaketitle

\section{Introduction}
%
%
%
%

Spectral imaging is a multidimensional data acquisition technology which captures a three-dimensional (3D) spectral data-cube $(x,y,\lambda)$ containing more information about the imaging scene than ordinary camera. Conventional spectral imaging, with point-to-point imaging mode, requires time scanning along either the spatial or wavelength axis~\cite{herrala1994imaging,green1998imaging}, resulting in limited application in some fields where high-speed imaging is needed. Hence, snapshot spectral imaging currently has aroused great interest, such as field-split imaging approach, computed tomography imaging approach and coded aperture spectral imaging approach~\cite{hagen2013review,sahoo2017single}. As a novel snapshot spectral imager, a spectral camera based on ghost imaging via sparsity constraints (GISC)~\cite{liu2016spectral} which realizes three-dimensional (3D) spectral imaging in a single shot through a spatial random phase modulator, can be one of the essential tools for high-speed spectral imaging in today's scientific research, industry and defense~\cite{gao2014single,donoho1992superresolution,morgan2003surface}.

In ghost imaging (GI), information of imaging  objects is extracted from the second-order correlation of fluctuating light fields. The performance of GI is theoretically related to the contrast $g^2$ of speckle patterns~\cite{Gong2010A}. Specifically, higher contrast of the speckle patterns may enable the enhancement in the signal-to-noise ratio (SNR) of the reconstructed images. Therefore,  customizing the speckle intensity statistics to generate super-Rayleigh speckles whose contrast is greater than 1 may improve the noise immunity of GISC spectral camera.

Generally, a universal spatial random phase modulator, whose phase is uniformly distributed over a range $[0, 2\pi]$ and each pixel is independent, generates Rayleigh speckles~\cite{goodman2007speckle}. There has aroused great interest in generating non-Rayleigh speckles for the last few years~\cite{cagigal2001experimental,apostol2005non,waller2012phase,
rodenburg2014experimental,amaral2015tailoring,li2016generation,kubota2010very,
fischer2015light,safari2017generation}.
Recently, Bromberg \emph{et al}. developed a simple method for producing non-Rayleigh speckles with a phase-only SLM to redistribute the intensity among the speckle grains~\cite{bromberg2014generating}. 
In this method, a phase-only SLM is illuminated by an expanded monochromatic laser beam, and the tailored speckle patterns are realized on the Fourier plane of the SLM. For GI, this method is suitable for GISC lidar~\cite{Zhao2012Ghost,chen2013ghost,Gong2016Three}, but not applicable to GISC spectral camera. Because in GISC spectra camera, the distance between the speckle detection plane and SLM plane needs to be flexibly adjusted according to the required system parameters, and the imaging scenarios of GISC spectral camera cover a broad spectral range. Hence, a universal and flexible method for generating super-Rayleigh speckles in GISC spectral camera needs to be investigated.

\begin{figure*}[!t]
\centering
\includegraphics[scale=0.6]{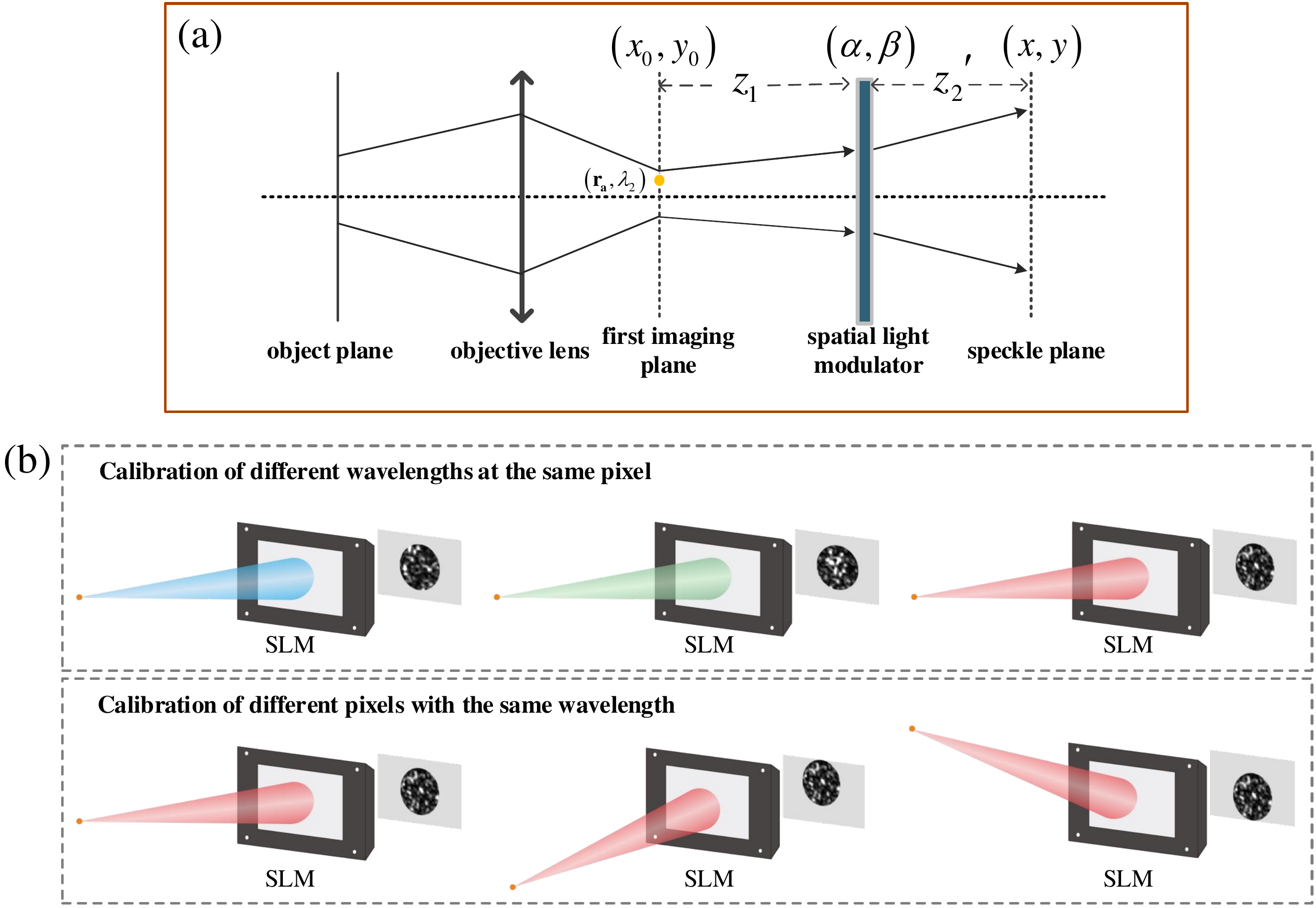}
\caption{\label{fig:system} GISC spectral camera. (a) The schematic of GISC spectral camera. (b) The simple sketch of the calibration process. From top  to bottom are: point sources with different wavelengths at the same pixel of field-of-view (FoV) illuminating the SLM to generate speckle patterns, point sources with the same wavelength  at different pixels of FoV illuminating the SLM to generate speckle patterns.}
\end{figure*}

In this work, based on the principle of reversibility of light, a flexible method for generating non-Rayleigh speckles with a phase-only SLM is demonstrated. Through the method, non-Rayleigh speckles can be realized within a range of axial distance behind the SLM by adaptively designing different phase matrices loaded on the SLM. Meanwhile, the effect of different system parameters in the formation of super-Rayleigh speckles is theoretically analyzed. Simulation and experimental results show that GISC spectral camera with super-Rayleigh modulator can significantly improve the quality of the reconstructed images, especially in the case of low SNR and low sampling rate.

The organization of this paper is as follows. In Section \uppercase\expandafter{\romannumeral2}, the schematic of GISC spectral camera is briefly introduced. Section \uppercase\expandafter{\romannumeral3 } compares the reconstruction accuracy using super-Rayleigh speckles, Rayleigh speckles and sub-Rayleigh speckles, indicating the superiority of using super-Rayleigh speckles in improving the reconstruction accuracy of imaging. The effect of different system parameters in the formation of super-Rayleigh speckles is also theoretically analyzed. Section \uppercase\expandafter{\romannumeral4} describes experimental setup and gives the experimental results, demonstrating the advantages of GISC spectral camera with super-Rayleigh modulator in enhancing the quality of reconstructed results. At last, some concluding remarks are drawn in Section \uppercase\expandafter{\romannumeral5}.

\section{System of GISC spectral camera}

GISC spectral camera is a snapshot spectral imager in which 3D spectral data-cube can be modulated into 2D speckle patterns by a spatial light modulator~\cite{liu2016spectral}. The  system is composed of four modules (see Fig. \ref{fig:system}(a)), (1) imaging module (an objective), which projects a scene onto the first imaging plane, (2) modulation module (a spatial light modulator), which modulates the light fields of different wavelengths and positions to generate different speckle patterns, (3) detection module (CCD), which records the speckle patterns on the speckle plane , (4) demodulation module, which recovers a target image via optimization algorithm.

\begin{figure*}[ht]
\centering
\includegraphics[scale=0.33]{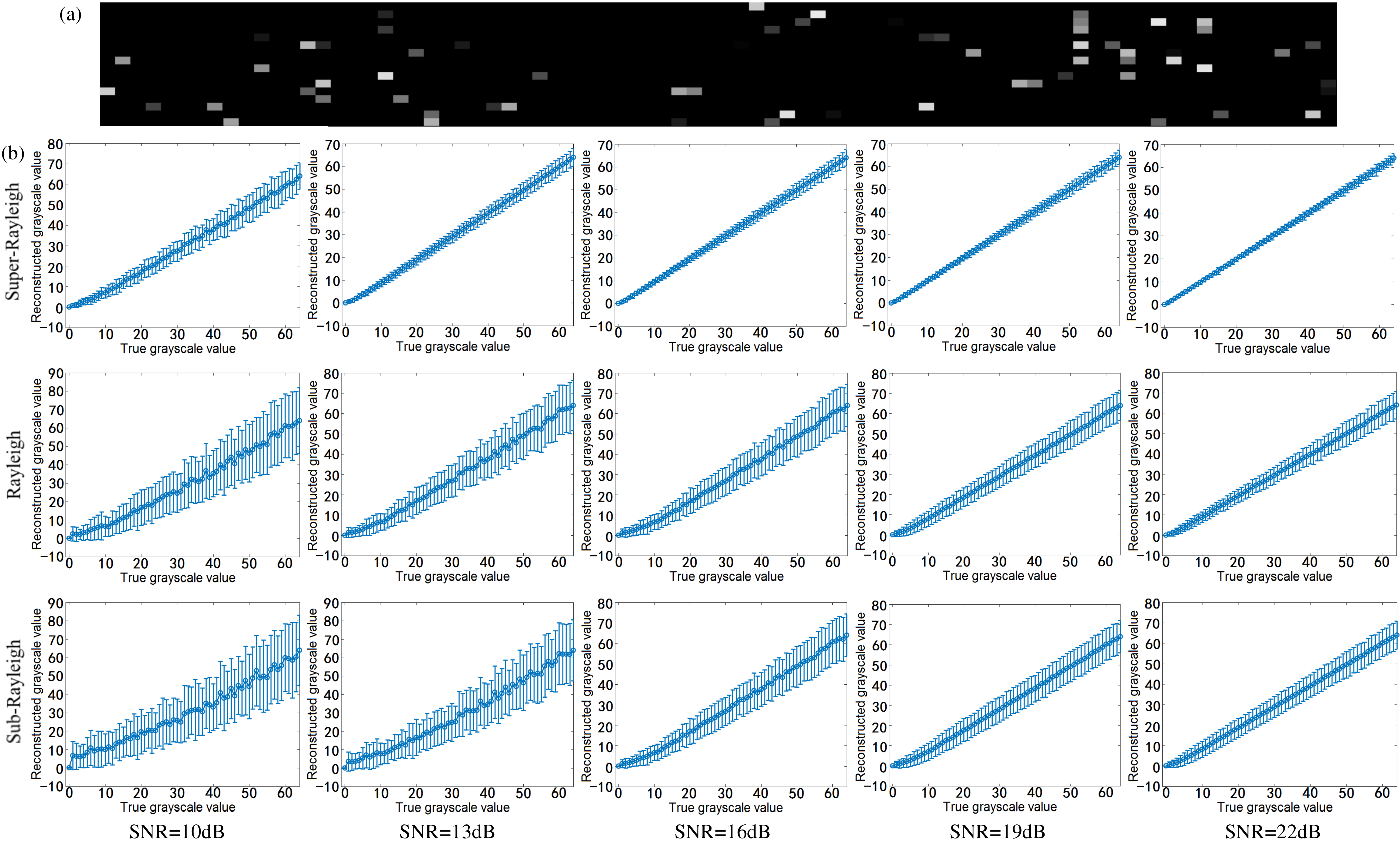}
\caption{The analysis of the reconstruction accuracy with different speckle patterns. (a) Original imaging scenario ($129\times129\times10$): $129\times129$ pixels, 10 spectral channels from 542 nm to 560 nm. (b) The dependence relation of the grayscale value of the reconstructed image with super-Rayleigh speckles on that of the original image under the sampling rate $C_R=50\%$ and different SNRs: SNR=10 dB, SNR=13 dB, SNR=16 dB, SNR=19 dB, SNR=22 dB. (c) The dependence relation of the grayscale value of the reconstructed image with Rayleigh speckles on that of the original image under the sampling rate $C_R=50\%$ and different  SNRs: SNR=10 dB, SNR=13 dB, SNR=16 dB, SNR=19 dB, SNR=22 dB. (d) The dependence relation of the grayscale value of the reconstructed image with sub-Rayleigh speckles on that of the original image under the sampling rate $C_R=50\%$ and different  SNRs: SNR=10 dB, SNR=13 dB, SNR=16 dB, SNR=19 dB, SNR=22 dB.}
\label{fig:grayscale}
\end{figure*}

The discrete model of the imaging process of GISC spectral camera is given by 
\begin{equation}\label{eq:a5}
\bm{Y}=\bm{AX+\omega},
\end{equation}
where $\bm{\omega}$ is additive noise, $\bm{A}$ is the measurement matrix which can be obtained by calibration process. The whole calibrating measurement can also be regarded as obtaining incoherent intensity impulse response function from different 3D spectral data-cube pixels within the field-of-view (FoV), as shown in Fig. \ref{fig:system}(b).

In the process of the calibration, the whole spectral band is divided into $L$ equispaced spectral channels, and the FoV is divided into $N$ pixels. Denote $p$th speckle intensity with wavelength $\lambda$ recorded by CCD as $I_p^\lambda$ ($M=l\times n$ pixels), and then reshape it into a column vector ${\bm{A}}_p^\lambda  = {\left( {A_{1,p}^\lambda ,A_{2,p}^\lambda , \cdots ,A_{M,p}^\lambda } \right)^T}$. The scene is denoted as a column vector ${{\bm{X}}_{L \times N}} = {\left( {{x_1},{x_2}, \cdots ,{x_{L \times N}}} \right)^T}$, and then the imaging model in matrix form can be expressed as 
\begin{equation}\label{eq:a6}
{y_q} = \sum\limits_{\lambda  = 1}^L {\sum\limits_{s = 1}^N {A_{q,s}^\lambda x_s^\lambda } }  + {\omega _q},
\end{equation}
where $y_q$ denotes the signal of the $q$th pixel on the CCD. The sampling rate ($C_R$) of the system is defined as ${C_R} = {M \mathord{\left/
 {\vphantom {M {\left( {L \times N} \right)}}} \right.
 \kern-\nulldelimiterspace} {\left( {L \times N} \right)}}$, where $M$ is the pixel numbers randomly chosen from the CCD which records the detection signal.
 
The 3D signal reconstruction is obtained by solving the following inverse problem with TV-RANK algorithm~\cite{shiyu2015hyperspectral}
\begin{equation}\label{eq:a7}
{\bm{X}} = \arg \mathop {min}\limits_{{\bm{X}} \ge 0} \left\| {{\bm{Y}} - {\bm{AX}}} \right\|_2^2 + {\mu _1}{\left\| {{\bm{\Phi X}}} \right\|_1} + {\mu _2}{\left\| {\bm{X}} \right\|_*},
\end{equation}
where ${\left\| {\bm{X}} \right\|_*}$ is nuclear norm, $\mu_1$, $\mu_2>0$ are the weight coefficients, ${\left\| {{\bm{\Phi X}}} \right\|_1}$ is the spatial total variation (TV).

\begin{figure*}[t]
\centering
\includegraphics[scale=0.62]{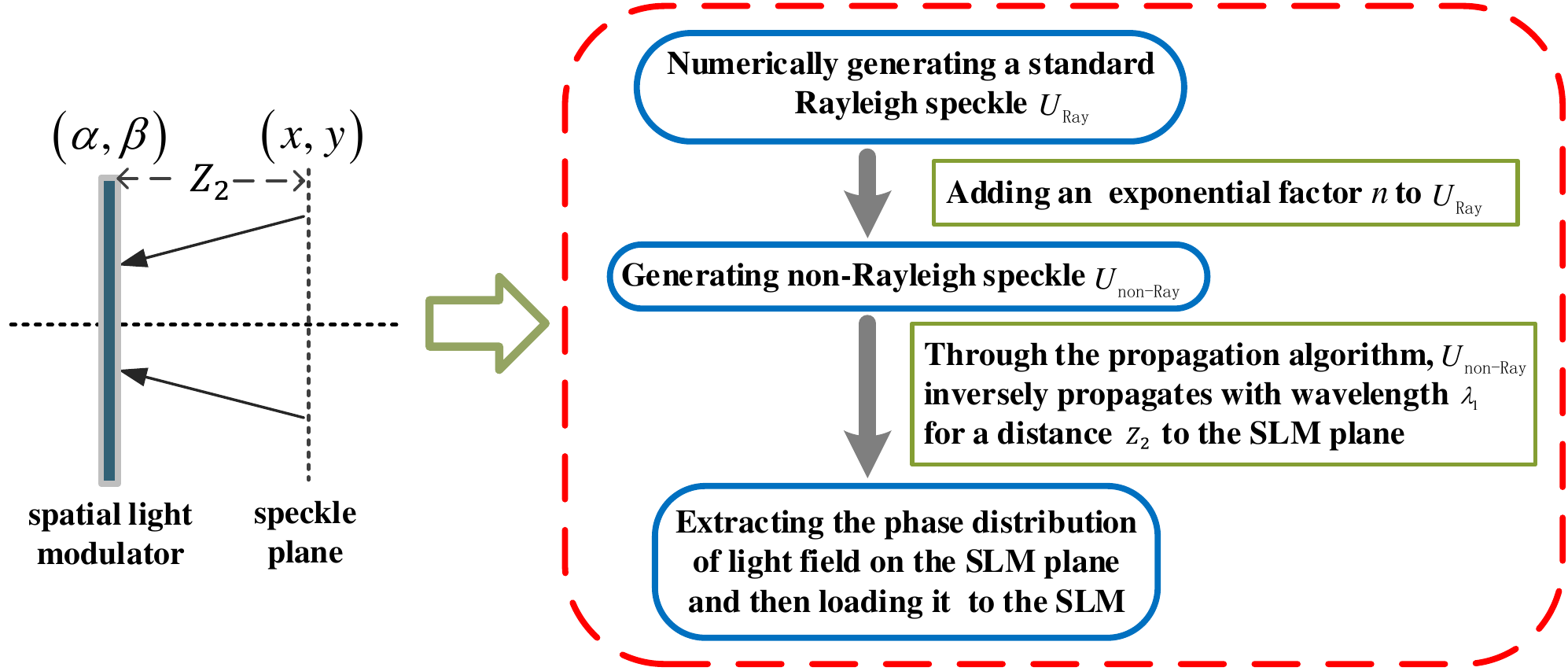}
\caption{\label{fig:speckle_generation} The process of generating a phase matrix loaded on SLM for realizing non-Rayleigh speckle patterns.}
\end{figure*}

\begin{figure*}[t]
\centering
\includegraphics[scale=0.56]{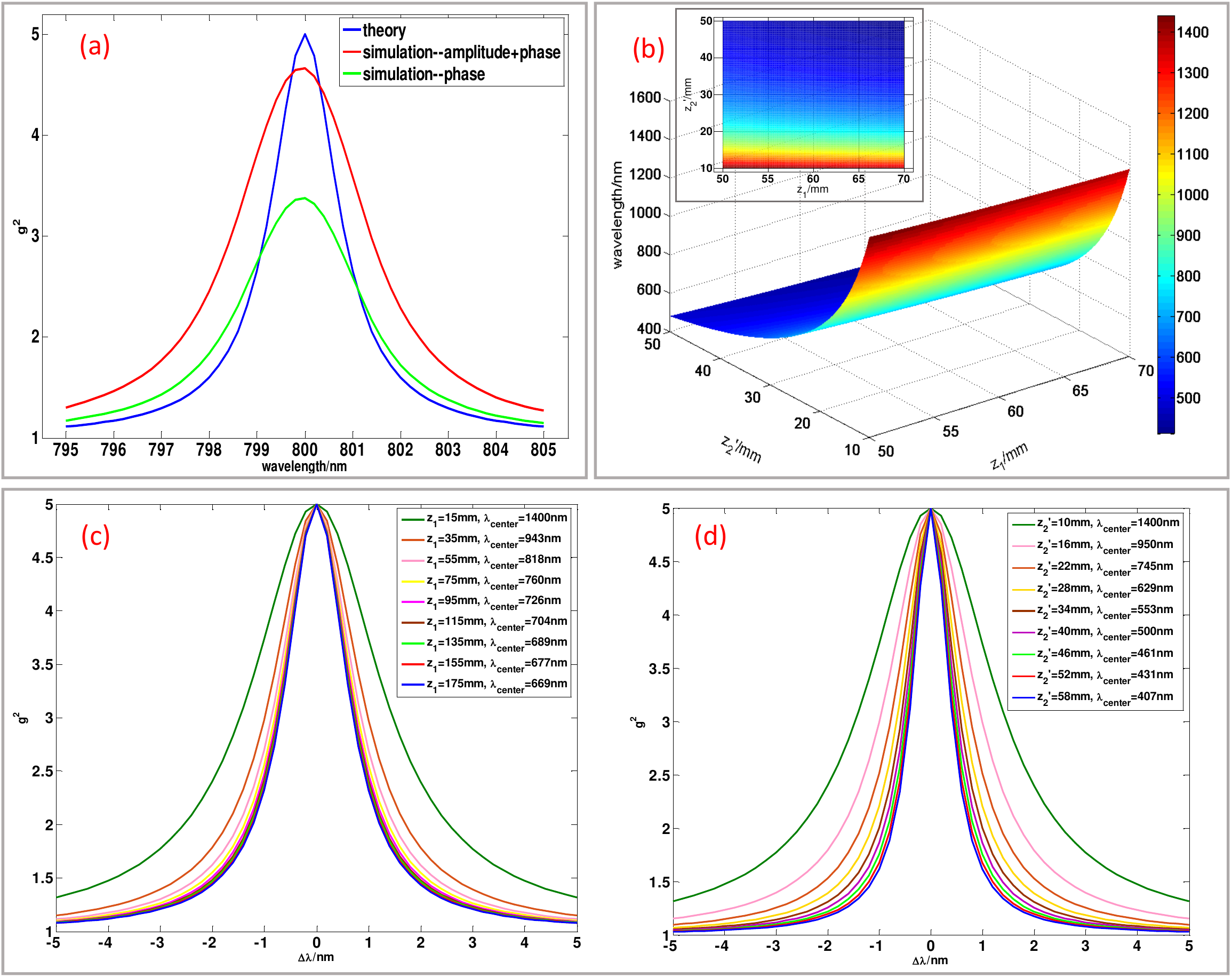}
\caption{Results of theoretical calculation and numerical simulation. (a) The contrast $g^{2}$ as a function of the wavelength with $z_1=60$ mm, ${z_2}'=20 $ mm. The red curve corresponds to the case when the matrix containing both the amplitude and phase information of $U(\bm{r_0},\lambda_1)$ is loaded onto the SLM and the green curve corresponds to the case when the matrix only containing the phase of $U(\bm{r_0},\lambda_1)$ is loaded onto the SLM. (b) The central wavelength $\lambda_{\rm{center}}$ for different $z_1$ and ${z_2}'$. (c) When ${z_2}'=20$ mm, the contrast $g_2$ varies with different wavelengths within $10 \rm{nm}$ spectral range at different values of $z_1$. (d) When ${z_1}=60$ mm, the contrast $g_2$ varies with different wavelengths within $10 \rm{nm}$ spectral range at different values of ${z_2}'$.}
\label{fig:theory-simulation}
\end{figure*}

\section{Theory and method}
\subsection{The analysis of the reconstruction accuracy with three types of speckle patterns}

To investigate the effect of super-Rayleigh speckles on the reconstruction performance, we evaluated and analyzed the reconstruction accuracy through simulation. As a comparison, the results of Rayleigh speckles and sub-Rayleigh speckles are also analyzed.

In the simulation, we use a spatial random phase modulator whose phase uniformly distributed over a range $[0,2\pi]$ and each pixel is independent to generate Rayleigh speckles, then add exponential factors 4 and 0.5 to Rayleigh speckles to respectively acquire super-Rayleigh speckles and sub-Rayleigh speckles~\cite{hang2015Ghost,Zhao2012Ghost}. According to the imaging model of GISC spectral camera, the FoV is divided into $129\times129$ pixels, the whole spectral band is divided into 10 equispaced spectral channels from 542 nm to 560 nm, and the sampling rate is selected as 50\%, then $m\times n$ measurement matrices $\bm{A}$ ($m=129\times 129\times 10\times 50\% $ and $n=129\times 129\times 10$) of three types of speckle patterns can be constructed. For each type of speckle patterns, we perform 40 independent trials by using 40 random images with $129\times129$ pixels and 10 spectral channels as the original targets $\bm{X}$. Each image contains 64 rectangular blocks ($l=8 \times 16$ pixels) with grayscale value of 1 to 64 and randomly placed in 10 spectral channels. Figure \ref{fig:grayscale}(a) shows one of these images. The detection signal $\bm{Y}$ is generated using Eq. (\ref{eq:a5}), and then the reconstructed image can be obtained by TV-RANK algorithm. For 40 reconstructed images corresponding to a specific type of measurement matrix, the mean value $\bar v^p$ and standard deviation $\sigma ^p$ of the reconstructed grayscale value of 40 rectangular blocks with the same true grayscale value $p$ can be calculated as follows,
\begin{equation}\label{eq:a8}
{\bar v^p} = \frac{1}{{s \times l}}\sum\limits_{i = 1}^s {\sum\limits_{j = 1}^l {v_{i,j}^p} },
\end{equation}
\begin{equation}\label{eq:a9}
{\sigma ^p} = \sqrt {\frac{1}{{s \times l}}\sum\limits_{i = 1}^s {\sum\limits_{j = 1}^l {{{\left( {v_{i,j}^p - {{\bar v}^p}} \right)}^2}} } },
\end{equation}
where $p=1,2,..,64$, $s=40$ is the number of random images, $l=8\times16$ is the  pixel numbers of each rectangular block and ${v_{i,j}^p}$ is the grayscale value at the $j$th pixel of the rectangular block with true grayscale value of $p$ in the $i$th reconstructed image.  

In Fig. \ref{fig:grayscale}(b), we plot $\bar v^p$ performance for each  speckle pattern as a function of the corresponding true grayscle value under different SNRs (in dB), where the length of each error bar is the calculated standard deviation $\sigma^p$. The SNR is defined as SNR=$10\lg( {{{\overline{y}}} \mathord{\left/
 {\vphantom {{{\overline{y}}} {{\sigma _{\omega}}}}} \right.
 \kern-\nulldelimiterspace} {{\sigma _{\omega}}}})$\cite{goodman2015statistical}, where $\overline{y} $ and $\sigma _{\omega} $  respectively denote  mean value of the detection signal $Y$ and standard deviation of the noise. 
 
The simulation results show that, for the whole SNR region under test, the error of the reconstructed result using super-Rayleigh speckles is significantly lower than that of Rayleigh speckles and sub-Rayleigh speckles. Hence, the utilization of super-Rayleigh speckles may have great advantages in improving the reconstruction accuracy of the imaging results.

\begin{figure*}[t]
\centering
\includegraphics[scale=0.2]{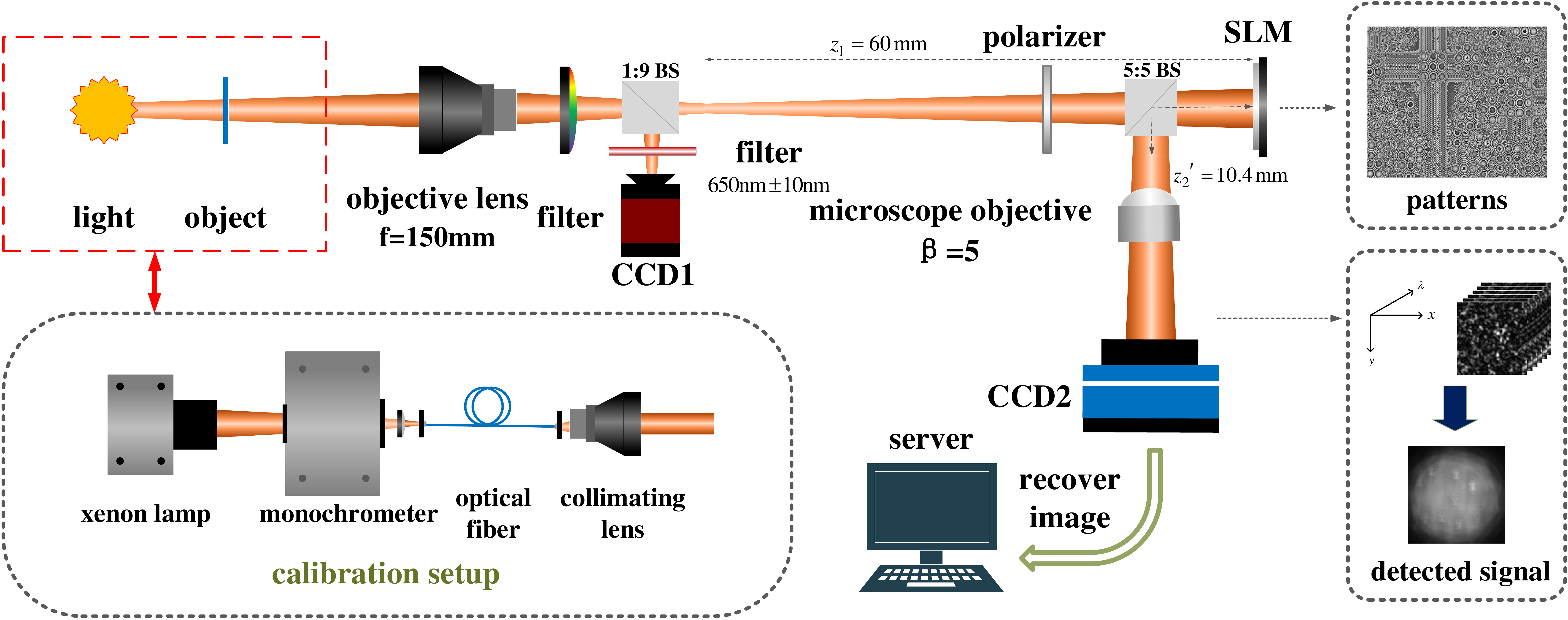}
\caption{The experimental setup. During the calibration, a phase matrix of desire is loaded onto the SLM and a calibration setup is put in front of the objective lens to acquire incoherent intensity impulse response functions of the system. After the calibration, an object is put on the object plane to obtain the detected signal which is the superposition of the speckles from different pixels and wavelengths of the object.}
\label{fig:experiment_setup}
\end{figure*}

\subsection{The influence of different system parameters of GISC spectral camera in the formation of super-Rayleigh speckle patterns}

The method for generating non-Rayleigh speckles proposed by Bromberg \emph{et al}.~\cite{bromberg2014generating} obtains non-Rayleigh speckle patterns on the Fourier plane of SLM. This method is not suitable for GISC spectral camera where  the distance between the speckle detection plane and SLM plane needs to be flexibly adjusted according to the required system parameters. Therefore, according to the imaging structure and principle of GISC spectral camera,  a universal and flexible method for generating non-Rayleigh speckles should be developed. Based on the principle of reversibility of light, a new phase matrix loaded on SLM which enables the higher order correlation of the scattered partial waves patterns~\cite{bromberg2014generating,zhang2015ghost,mandel1995optical} can be obtained through the steps in Fig. \ref{fig:speckle_generation}. 

Firstly, an exponential factor $n$ is added to Rayleigh speckle field ${U_{\rm{Ray}}}\left( {{{\bm{r_{1}}}}} \right)$ to obtain non-Rayleigh speckle field $W\left( {{\bm{r_{1}}}} \right) = {\left( {{U_{\rm{Ray}}}\left( {{\bm{r_{1}}}} \right)} \right)^n}$~\cite{bromberg2014generating}, where ${\bm{r_{1}}}$ represents a two-dimensional vector on the $xy$ plane. Then the light field  $W\left( {{\bm{r_{1}}}} \right)$  with wavelength $\lambda_{1}$ inversely propagates for a distance $z_2$ to the SLM plane $\alpha\beta$, and the light field on the $\alpha\beta$ plane can be denoted as
\setlength{\arraycolsep}{0.0em}
\begin{equation}\label{eq:a1}
\begin{aligned}
U\left( {{{\bm{r_{0}}}},{\lambda _1}} \right) =& \frac{j}{{{\lambda _1}{z_2}}}\exp \left( { - j\frac{{2\pi }}{{{\lambda _1}}}{z_2}} \right)\int\limits_{ - \infty }^\infty  W\left( {{{\bm{r_{1}}}}} \right)\\
&\times \exp \left( { - j\frac{\pi }{{{\lambda _1}{z_2}}}{{\left( {{{\bm{r_{0}}}} - {{\bm{r_{1}}}}} \right)}^2}} \right) d{{\bm{r_{1}}}},
\end{aligned}
\end{equation}
where ${\bm{r_{0}}}$ represents a two-dimensional vector on the $\alpha\beta$ plane. Finally, the phase matrix can be obtained by extracting the phase distribution of the light field $U\left( {{{\bm{r_{0}}}},{\lambda _1}} \right)$.

To investigate the role of system parameters of GISC spectral camera in the formation of super-Rayleigh speckles, we assume that a monochromatic point source at pixel $\bm{r_{a}}$ with wavelength $\lambda_{2}$ on the $x_{0}y_{0}$ plane shown in Fig. \ref{fig:system}(a) propagates to the speckle plane at a distance of ${z_{2}}'$ behind the SLM. Here, for theoretical derivation, the transmittance function of the SLM is defined as $T\left( {{\bm{r_0}},\lambda } \right) = U\left( {{\bm{r_0}},\lambda } \right)$. Then according to Fresnel diffraction theorem, the intensity distribution on the speckle plane can finally calculated as
\setlength{\arraycolsep}{0.0em}
\begin{equation}
\begin{aligned}\label{eq:a3}
&I\left( {\bm{r_1},{\lambda _2}} \right) = {\chi ^2}\iint {W}\left( \bm{\xi } \right){W^*}\left( \bm{\xi '} \right)\exp \left( {\frac{{ - j\pi }}{{{\lambda _1}{z_2}\gamma }}} \right.\left( {\left( {\gamma  + \frac{{{z_2}}}{{{z_1}}}} \right)} \right.\\
&\left. {\left. { \times \left( {\bm{\xi ^2} - {\bm{\xi '}^2}} \right) - 2\bm{r_1}\gamma \left( {\bm{\xi}  -\bm{ \xi '}} \right)} \right) - 2\frac{{{z_2}}}{{{z_1}}}\bm{r_a}\left( {\bm{\xi}  - \bm{\xi '}} \right)} \right)\rm{d}\bm{\xi} \rm{d}\bm{\xi '},
\end{aligned}
\vspace{8pt}
\end{equation}
where $\beta  = {{{\lambda _1}} \mathord{\left/ {\vphantom {{{\lambda _1}} {{\lambda _2}}}} \right. \kern-\nulldelimiterspace} {{\lambda _2}}}$, $\gamma  = {{{z_2}} \mathord{\left/ {\vphantom {{{z_2}} {{z_2}^\prime }}} \right.
 \kern-\nulldelimiterspace} {{z_2}^\prime }}$, $W\left( {{\bm{\xi}}} \right) = {\left( {{U_{\rm{Ray}}}\left( {{\bm{\xi}}} \right)} \right)^n}$, $\chi {\rm{ = }}{{\beta \gamma {z_1}} \mathord{\left/
 {\vphantom {{\beta \gamma {z_1}} {\left( {\left( {\beta \gamma  - 1} \right){z_1} + \beta {z_2}} \right)}}} \right. \kern-\nulldelimiterspace} {\left( {\left( {\beta \gamma  - 1} \right){z_1} + \beta {z_2}} \right)}}$. Here for convenience, $n \in {N^*}$ is considered. The contrast of the speckle patterns is defined as ${g^2} = {{\left\langle {I{{\left( {{{\bm{r}}_{\bm{1}}},{\lambda _2}} \right)}^2}} \right\rangle } \mathord{\left/
 {\vphantom {{\left\langle {I{{\left( {{{\bm{r}}_{\bm{1}}},{\lambda _2}} \right)}^2}} \right\rangle } {{{\left\langle {I\left( {{{\bf{r}}_{\bf{1}}},{\lambda _2}} \right)} \right\rangle }^2}}}} \right.
 \kern-\nulldelimiterspace} {{{\left\langle {I\left( {{{\bm{r}}_{\bm{1}}},{\lambda _2}} \right)} \right\rangle }^2}}} - 1$, where $\left\langle...\right\rangle$ denotes ensemble averaging. Assuming that the autocorrelation of Rayleigh speckle ${U_{\rm{Ray}}}\left( {{{\bm{r_{1}}}}} \right)$ obeys Gaussian distribution $ \left\langle {{U_{\rm{Ray}}}\left( {\bm{r_{1}}} \right)U_{\rm{Ray}}^*\left( {{\bm{r_{1} '}}} \right)} \right\rangle  =  {{\exp \left\{ { - {{{{\left( {{\bm{r_{1} }} - {\bm{r_{1} '}}} \right)}^2}} \mathord{\left/
 {\vphantom {{{{\left( {{\bm{r_{1} }} - {\bm{r_{1} '}}} \right)}^2}} {2{\sigma ^2}}}} \right.
 \kern-\nulldelimiterspace} {2{\sigma ^2}}}} \right\}} \mathord{\left/
 {\vphantom {{\exp \left\{ { - {{{{\left( {{\bm{r_{1} }} - {\bm{r_{1} '}}} \right)}^2}} \mathord{\left/
 {\vphantom {{{{\left( {{\bm{r_{1} }} - {\bm{r_{1} '}}} \right)}^2}} {2{\sigma ^2}}}} \right.
 \kern-\nulldelimiterspace} {2{\sigma ^2}}}} \right\}} {\sqrt {2\pi \sigma } }}} \right.
 \kern-\nulldelimiterspace} {\sqrt {2\pi \sigma } }}$ where $\sigma$ is the standard deviation, and substituting Eq. (\ref{eq:a3}) into the expression of contrast, ${g^2}$ can be ultimately expressed as 
\setlength{\arraycolsep}{0.0em}
\begin{equation}\label{eq:a4}
{g^2} = 1 + \sum\limits_{k = 1}^{n - 1}\frac{{{{ {{{\left( {C_n^k} \right)}^4}{{\left( {k!\left( {n - k} \right)!} \right)}^2}} } \mathord{\left/
 {\vphantom {{\sum\limits_{k = 1}^{n - 1} {{{\left( {C_n^k} \right)}^4}{{\left( {k!\left( {n - k} \right)!} \right)}^2}} } {{{\left( {n!} \right)}^2}}}} \right.
 \kern-\nulldelimiterspace} {{{\left( {n!} \right)}^2}}}}}{{1 + {{k\left( {n - k} \right){z_2}^2{{\left( {{\lambda _1} - \frac{1}{{\left( {\gamma  + \tau } \right)}}{\lambda _2}} \right)}^2}} \mathord{\left/
 {\vphantom {{k\left( {n - k} \right){z_2}^2{{\left( {{\lambda _1} - \frac{1}{{\left( {\gamma  + \tau } \right)}}{\lambda _2}} \right)}^2}} {{\pi ^2}{\sigma ^4}}}} \right.
 \kern-\nulldelimiterspace} {{\pi ^2}{\sigma ^4}}}}},
 \vspace{8pt}
\end{equation} 
where $\tau  = {{{z_2}} \mathord{\left/ {\vphantom {{{z_2}} {{z_1}}}} \right.\kern-\nulldelimiterspace} {{z_1}}}$ (see appendix for details). From Eq. (\ref{eq:a4}), when ${\lambda _2} = \left( {\gamma + \tau } \right){\lambda _1}$, the maximum value of $g^{2}$ is given by ${g^2}_{\rm{max}} =1 + {{\sum\limits_{k = 1}^{n - 1} {{{\left( {C_n^k} \right)}^4}{{\left( {k!\left( {n - k} \right)!} \right)}^2}} } \mathord{\left/
 {\vphantom {{\sum\limits_{k = 1}^{n - 1} {{{\left( {C_n^k} \right)}^4}{{\left( {k!\left( {n - k} \right)!} \right)}^2}} } {{{\left( {n!} \right)}^2}}}} \right.
 \kern-\nulldelimiterspace} {{{\left( {n!} \right)}^2}}}$. Figure \ref{fig:theory-simulation}(a) (blue line) shows the contrast ${g^2}$ of the speckle patterns as a function of the wavelength with certain parameters $z_1$, $z_2$, $z_2^{'}$, $\lambda_1$ and $n$, showing that the contrast of super-Rayleigh speckles is dependent on the wavelength and decreases as the wavelength $\lambda_2$ gradually deviates from the central wavelength $\lambda_{\rm{center}}$ which is defined as the wavelength when $g^2$ is maximized.

In order to verify the result of the theoretical derivation, the process of generating super-Rayleigh in GISC spectral imager is numerically simulated by propagation algorithm. Figure \ref{fig:theory-simulation} shows the results of theoretical calculation and numerical simulation when $n=2$, $z_2=20$ mm, $\lambda_1=600$ nm, $\sigma=2.55$ $\rm{\mu m}$. As shown in Fig. \ref{fig:theory-simulation}(a), the dependence of $g^{2}$ on wavelength obtained through simulation is basically consistent with the theoretical derivation.

From Figs. \ref{fig:theory-simulation}(b)-\ref{fig:theory-simulation}(d), we can see that system parameters $z_1$ and ${z_2}'$ affect the dependence relation of $g^2$ on the central wavelength $\lambda_{\rm{center}}$. In addition, $\lambda_{\rm{center}}$ gradually shifts to shorter waves and the spectral range within which the speckle fields maintain super-Rayleigh distribution gradually narrows as $z_1$ or ${z_2}'$ increases.

\begin{figure}[t]
\centering
\includegraphics[scale=0.28]{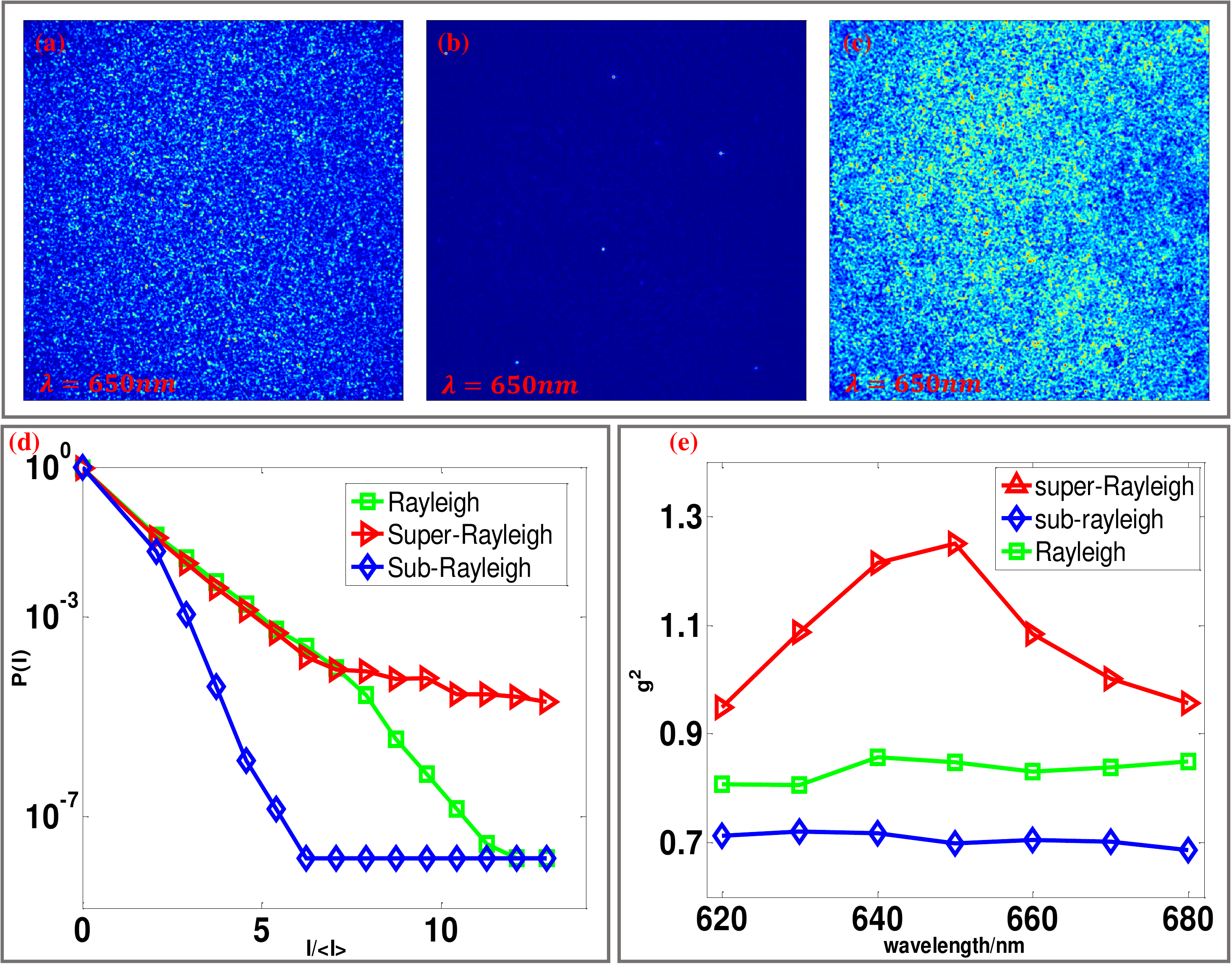}
\caption{Experimental generation of speckles. (a)-(c) Images of Rayleigh speckle patterns, super-Rayleigh speckle patterns and sub-Rayleigh speckle patterns with the wavelength 650nm. (d) The corresponding intensity probability distribution function. (e) The normalized second order correlation function ${g_2}$ from the wavelength 620 nm to 680 nm.}
\label{fig:experiment_speckle_result}
\end{figure}

\section{Experimental methods}
\subsection{Experimental setup}

In the experiment, the phase matrix loaded onto the SLM is obtained through the method described in the previous section. The parameters of generating phase matrix are set as $\lambda_1=650$ nm, $z_2=9$ mm, $n=8,0.3$, and the parameters of the experimental system are set as $z_1=60$ mm, ${z_2}'=10.4$ mm. The central wavelength $\lambda_{\rm{center}}$ is about 650 nm according to Eq. (\ref{eq:a4}). 

Figure \ref{fig:experiment_setup} shows the experimental setup. The objective lens with focal length of $f=150$ mm projects the unknown target image onto the first imaging plane. A $620\sim680$ nm band pass filter located behind the objective lens (Tamron AF70-300 \rm{mm} f/4-5.6) ensures that only the spectral data-cube corresponding to $620\sim680 $ nm band is measured by the system. A beam splitter with 1:9 splittering ratio in front of the first imaging plane splits the light field into two paths, and a surveillance camera CCD1 (AVT Sting F-504C owning 3.45 $\rm{\mu m}$ $\times$ 3.45 $\rm{\mu m}$) in one path records the conventional image as a reference. The other path of the light field passes through a polarizer and a beam splitter with 5:5 splittering ratio. The incident polarisation for phase only modulation is along the long display axis of SLM. Then the transmitted light field illuminates a phase-only reflective SLM (HOLOEYE PLUTO-VIS, resolution: $1920 \times 1080$ pixels, frame rate: 60 $\rm{Hz}$, pixel pitch: 8 $\rm{\mu m}$ $\times$ 8 $\rm{\mu m}$, fill factor: 87\%). According to the system parameters set above, the SLM is placed at $z_1=60$ \rm{mm} behind the first imaging plane. An microscope objective whose magnification is $\beta=5$ is placed at the position that makes its object surface (speckle plane) locate at ${z_2'}=10.4$ \rm{mm} behind the SLM. Then CCD2 (Apogee with 13 $\mathrm{\mu m}$ $\times$ 13 $\mathrm{\mu m}$) records the intensity  distribution of the amplified speckle fields in a single exposure.

Similar to the GISC spectral camera, the calibration setup of the experiment is shown in the gray dotted box at the bottom of Fig. \ref{fig:experiment_setup}. A xenon lamp produces thermal light to illuminate the entrance of a monochromator (WDG30-Z). Then quasi-monochromatic point light source is generated through coupling the quasi-monochromatic light field from the exit of the monochromator into an optical fiber with a diameter of 20 $\rm{\mu m}$. We put the output end of the optical fiber in the equivalent plane locating at the focal plane of a collimating lens (Olympus M.ZUIKO AF40-150 mm). And then the objective lens collects the parallel light field from the collimating lens~\cite{wu2014snapshot}. During the calibration, the imaging plane of the objective lens (the first imaging plane) is divided into $151 \times 151$ pixels, and the number of spectral channels is from 620 \rm{nm} to 680 \rm{nm} at interval of 10 \rm{nm}.

\begin{figure}[t]
\centering
\includegraphics[scale=0.26]{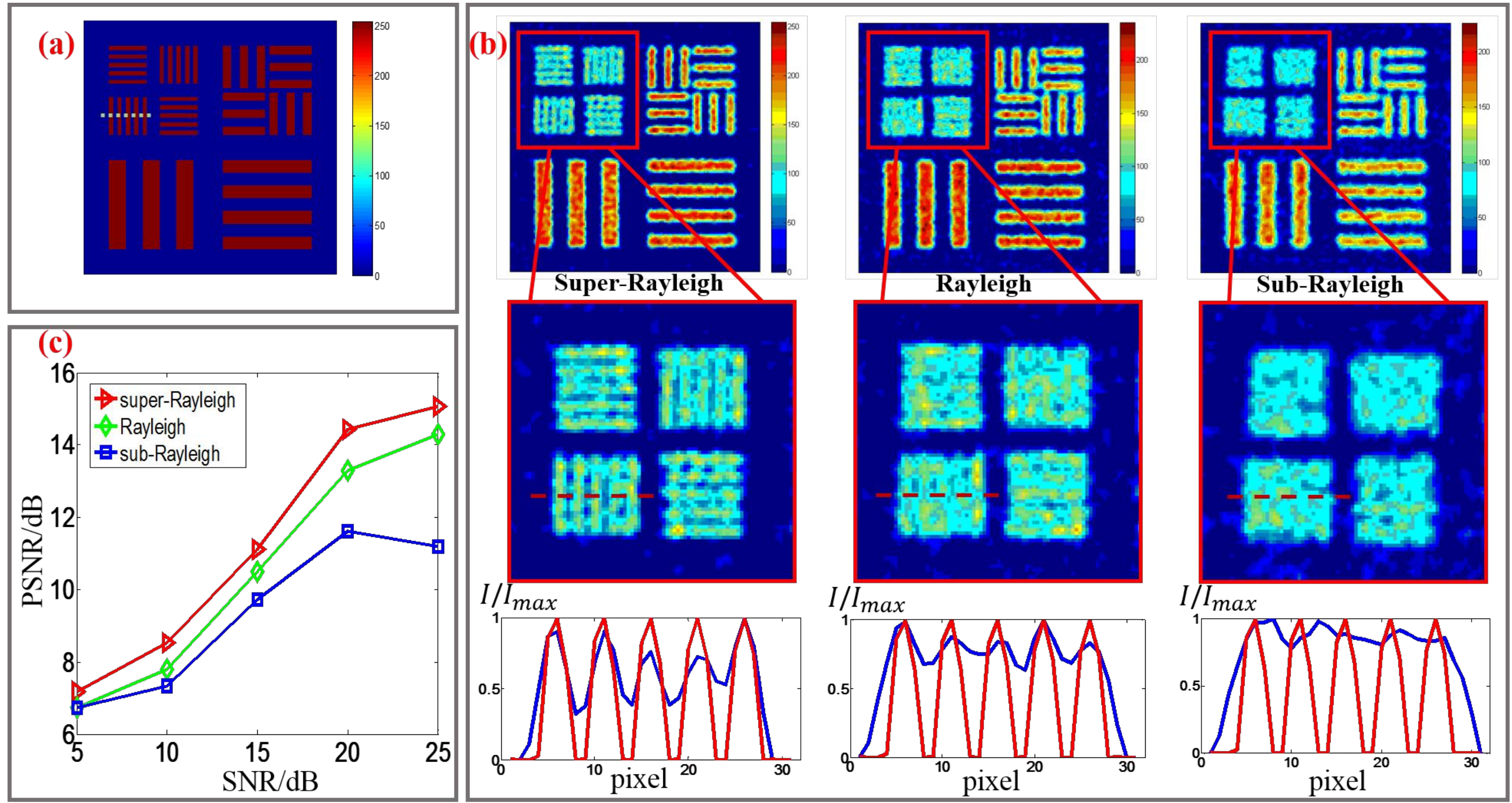}
\caption{Simulated imaging results. (a) Original image with pixels $151 \times 151$. (b) From left to right, the reconstructed images at the wavelength 650 nm obtained using super-Rayleigh speckles, Rayleigh speckles and sub-Rayleigh speckles respectively. The curves below are cross-sections of the original image (red line) and reconstructed image (blue line) obtained by these three types of speckles. The reconstructed image obtained by using super-Rayleigh speckles can distinguish the narrowest line pair, which is better than the other two types of speckles. (c) The computed peak signal to noise ratio (PSNR) of the reconstructed image obtained by using super-Rayleigh speckles (red triangles) is higher than Rayleigh speckles (green diamonds) and sub-Rayleigh speckles (blue squares).}
\label{fig:simulation_result}
\end{figure}
\begin{figure}[t]
\centering
\includegraphics[scale=0.38]{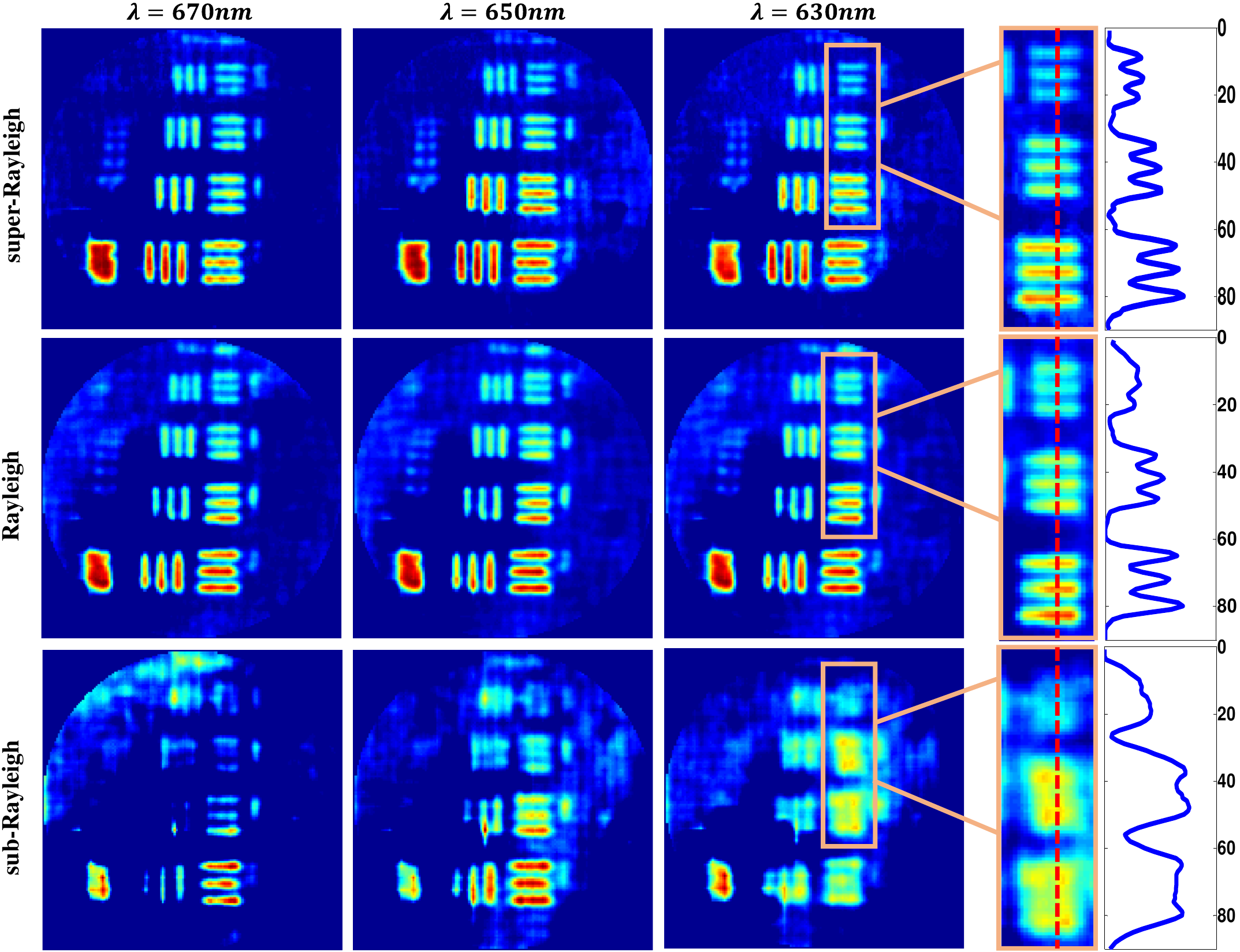}
\caption{Experimental imaging results of a resolution target at wavelengths of 630 nm, 650 nm and 670 nm. The reconstructed images and the resolution analysis curve on the right show that the resolution of the imaging results using super-Rayleigh speckles are better than that of Rayleigh speckles and sub-Rayleigh speckles.}
\label{fig:experiment_imaging_result_resolution}
\end{figure}

\begin{figure}[t]
\centering
\includegraphics[scale=0.23]{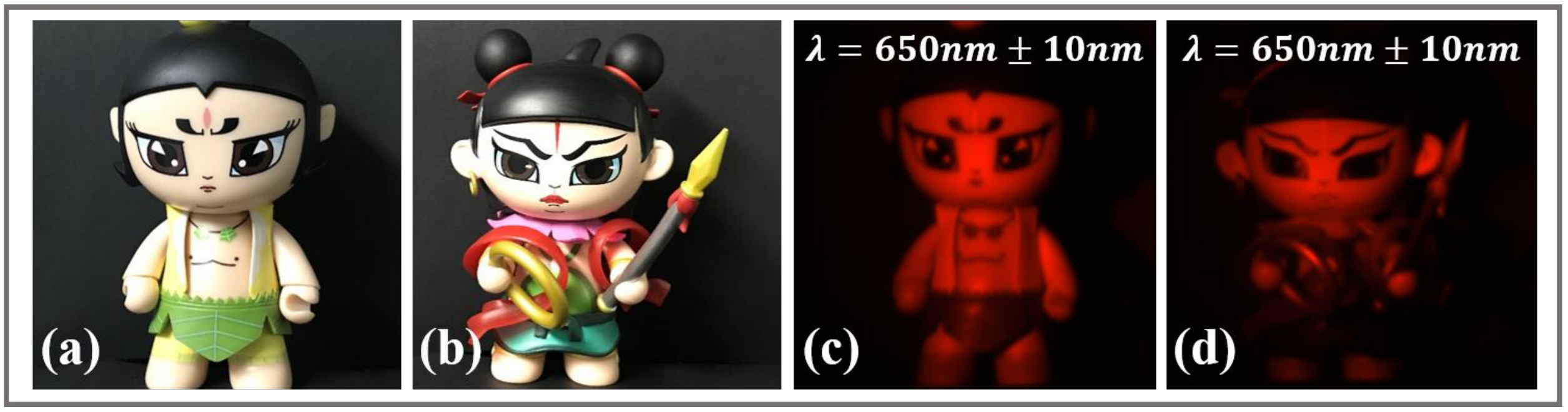}
\caption{(a)-(b) Physical image of the experimental object obtained by the conventional camera. (c)-(d) The images obtained by the CCD1.}
\label{fig:experiment_imaging_ccd}
\end{figure}

\begin{figure}[t]
\centering
\includegraphics[scale=0.323]{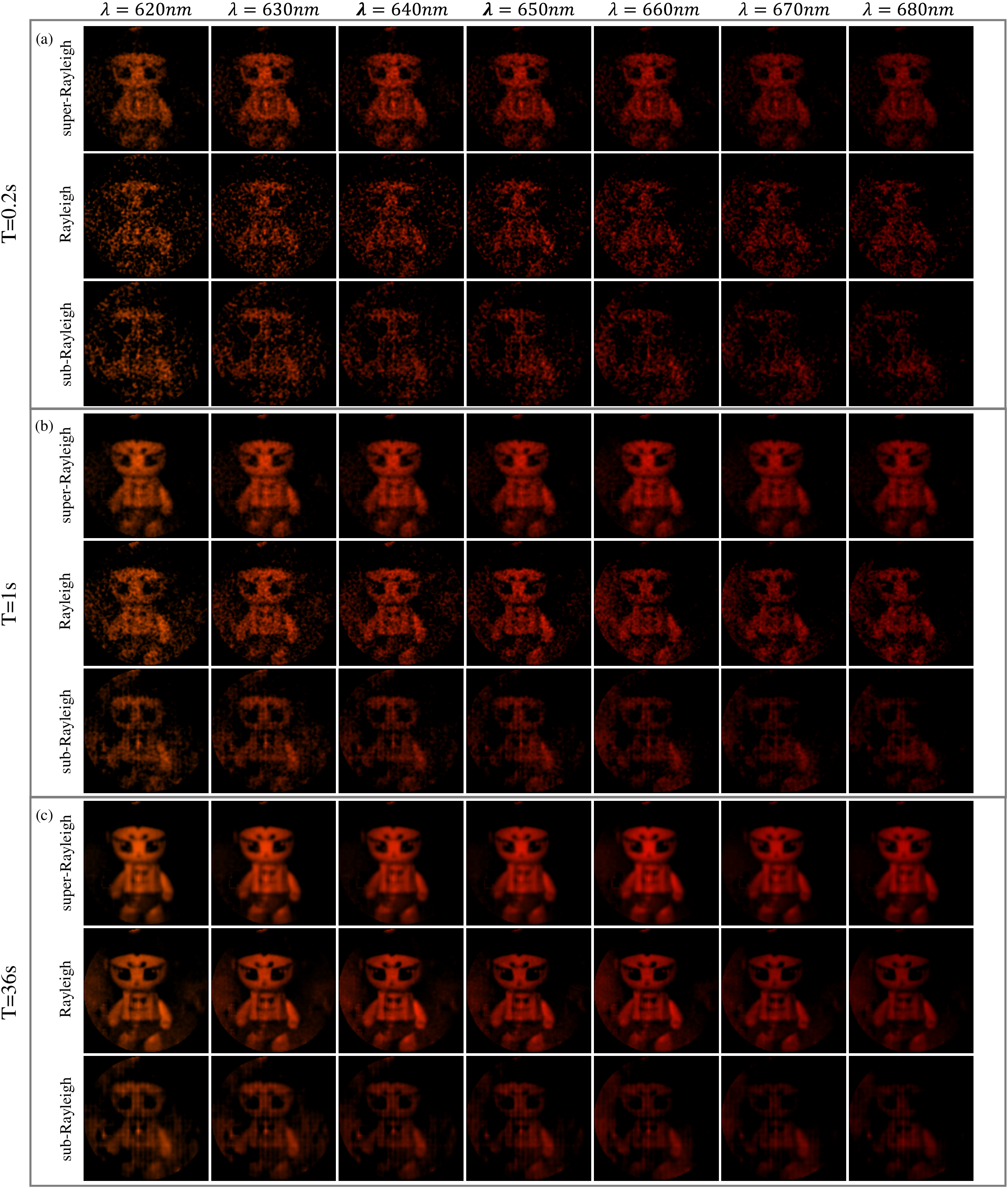}
\caption{Experimental imaging results of a toll with different exposure times. From top to bottom, the sampling rate is $50\%$ and the SNR of the detected signal is gradually increased. In the case of low SNR, the imaging results using super-Rayleigh speckles are significantly better than the other two speckles. }
\label{fig:experiment_imaging_result_huluwa}
\end{figure}

\begin{figure}[t]
\centering
\includegraphics[scale=0.323]{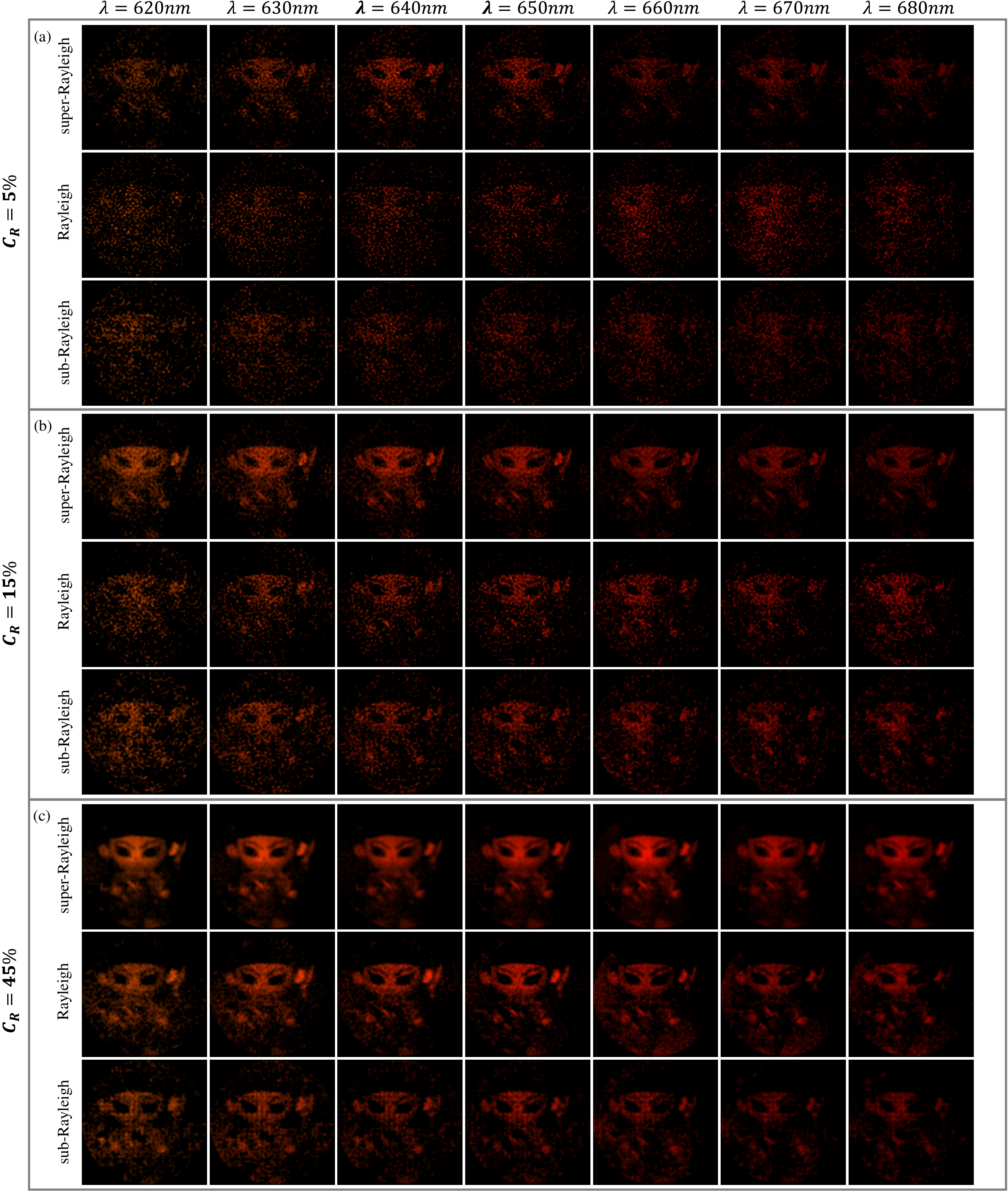}
\caption{Experimental imaging results of a toll with different sampling rates. From top to bottom, the exposure time is 36 seconds and the sampling rate ($C_R$) of the image is gradually increased. In the case of low sampling rate, the imaging results using super-Rayleigh speckles are significantly better than the other two speckles.}
\label{fig:experiment_imaging_result_nazha}
\end{figure}

\subsection{Experimental results}
In the calibration process, the measurement matrices of Rayleigh distribution and non-Rayleigh distribution can be acquired. Examples of experimental speckle patterns with different intensity statistics obtained by loading the phase matrices corresponding to different intensity distributions onto the SLM are shown in Figs. \ref{fig:experiment_speckle_result}(a)-\ref{fig:experiment_speckle_result}(c). Figures \ref{fig:experiment_speckle_result}(b) and \ref{fig:experiment_speckle_result}(c) illustrate that the speckle patterns of non-Rayleigh distribution is different from Rayleigh speckles shown in Fig. \ref{fig:experiment_speckle_result}(a). Indeed, the intensity probability distribution of non-Rayleigh speckles in Fig. \ref{fig:experiment_speckle_result}(d) [red triangles and blue squares] does not satisfy the negative exponential distribution [green diamonds]. Figure \ref{fig:experiment_speckle_result}(e) shows the contrast $g^2$ of different speckle patterns. It can be seen that $g^2$ of super-Rayleigh speckle field (red triangles) varies from 0.95 to 1.25 within the wavelength range of $620\sim680$ nm, which is higher than Rayleigh and sub-Rayleigh speckle patterns (green squares and blue diamonds). Therefore, the speckles can still maintain its super-Rayleigh statistics for a relatively wide spectral bandwidth although the calculation of $g^2$ has been only carried out for a specific wavelength.

We utilize the measurement matrix obtained by experiment for numerical simulation to quantitatively analyze the reconstructed results with different speckle patterns. A resolution target with pixels 151 $\times$ 151, as shown in Fig. \ref{fig:simulation_result}(a), is used in the numerical simulation. The wavelength and the SNR selected in this numerical simulation are respectively 650 nm and 20 dB. Figure \ref{fig:simulation_result}(b) shows the reconstructed results of the resolution target using three different types of speckles at $50\%$ sampling rate. Figure \ref{fig:simulation_result}(c) displays the peak signal-to-noise ratio (PSNR) of the reconstructed images. This figure clearly shows that, compared to Rayleigh speckles and sub-Rayleigh speckles, imaging with super-Rayleigh speckles enables higher resolution and SNR of the reconstructed images.

After the calibration process, the calibration setup shown in the gray box of Fig. \ref{fig:experiment_setup} is replaced by a real object for imaging experiments. Firstly, we used a transmissive resolution target to analyze the resolution of the reconstructed image. Figure \ref{fig:experiment_imaging_result_resolution} shows the reconstructed images at the wavelength 630 \rm{nm}, 650 \rm{nm}, and 670 \rm{nm}. From the experimental results in Fig. \ref{fig:experiment_imaging_result_resolution}, it can be seen that the resolution of the reconstructed images using super-Rayleigh speckles are better than Rayleigh speckles and sub-Rayleigh speckles, which is consistent with the above simulation results, thus further illustrating that the super-Rayleigh speckles is conductive to improving the image quality in practical application.

Afterwards, multispectral dolls, which are illuminated with thermal light, were utilized as imaging objects. Figures \ref{fig:experiment_imaging_ccd}(a) and \ref{fig:experiment_imaging_ccd}(b) are the results acquired by a conventional camera and Figs. \ref{fig:experiment_imaging_ccd}(c) and \ref{fig:experiment_imaging_ccd}(d) are the results detected by CCD1 of the experimental setup. In order to obtain the detection signal of Rayleigh speckles and non-Rayleigh speckles under different SNR, we select a series of different exposure times of 0.2 seconds, 1 seconds, 36 seconds when performing imaging. The results of the reconstructed images through optimized reconstruction algorithm are shown in Fig. \ref{fig:experiment_imaging_result_huluwa}. 
As expected, the quality of the reconstructed images using super-Rayleigh speckles is significantly better than Rayleigh speckles and sub-Rayleigh speckles, especially at low SNR. Figure \ref{fig:experiment_imaging_result_nazha} shows the reconstruction results at different sampling rates of 5\%, 15\% and 45\%. From Fig. \ref{fig:experiment_imaging_result_nazha}, it is clearly seen that the quality of the reconstructed images using super-Rayleigh speckles is better than the other two speckles especially at low sampling rate.

\begin{figure}[t]
\centering
\includegraphics[scale=0.245]{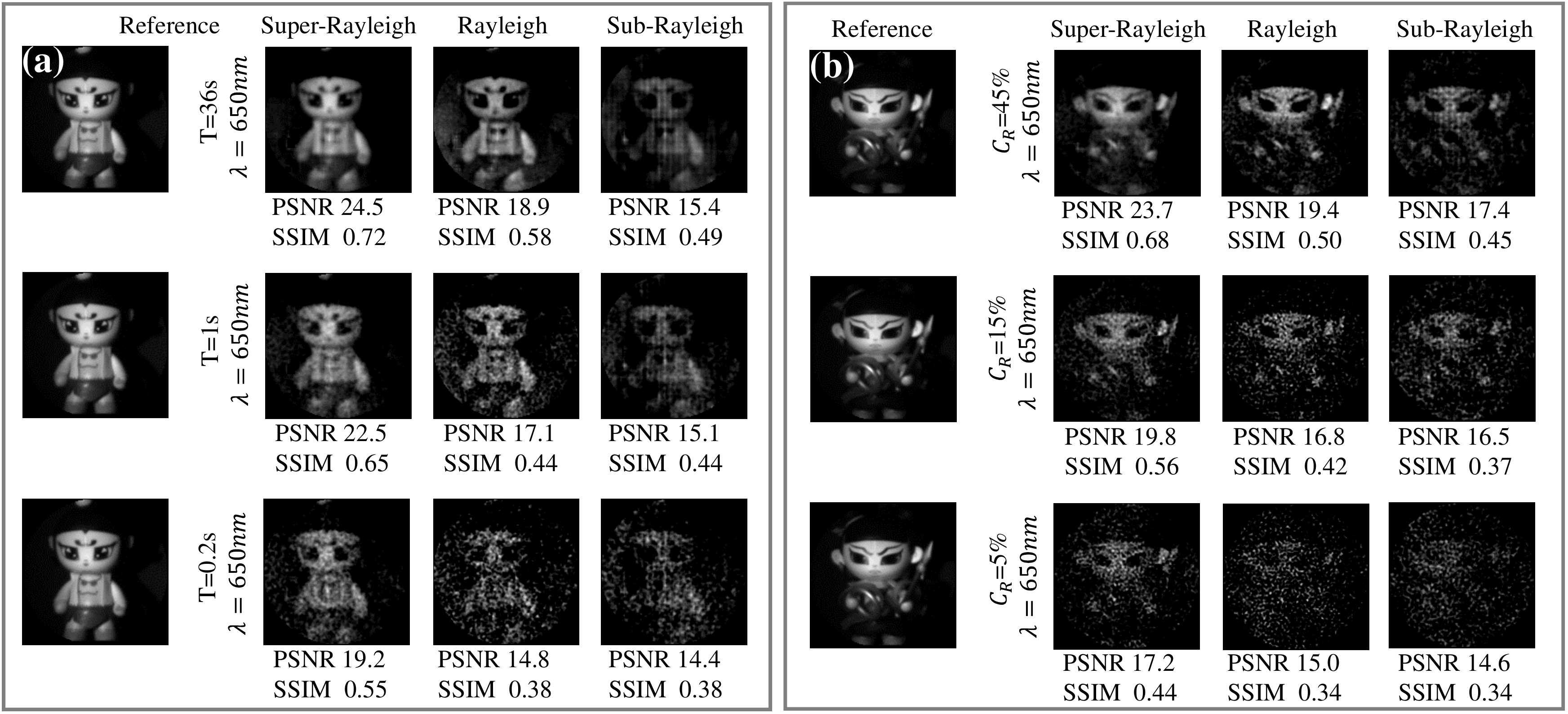}
\caption{The analysis of the peak signal-to-noise ratio (PSNR) and structural similarity index (SSIM) about the tolls with different exposure times and sampling rates. (a) The PSNR and SSIM of the reconstructed images obtained with different exposure times at the sampling rate of $50\%$ , from the calculation results, the quality of the images using super-Rayleigh speckles is better than the other two. (b) The PSNR and SSIM of the reconstructed images obtained with different sampling rates at the exposure time of 36 seconds, from the calculation results, the quality of the images using super-Rayleigh speckles is better than the other two.}
\label{fig:experiment_imaging_result_analyze}
\end{figure}

Moreover, in order to quantitatively analyze the quality of the reconstructed images, the peak signal-to-noise ratio (PSNR) and structural similarity index (SSIM)~\cite{wang2004image,hore2010image} of the reconstructed images using three different types of speckle patterns are calculated, as shown in Fig. \ref{fig:experiment_imaging_result_analyze}. From the results, the reconstructed images using super-Rayleigh speckles exhibit superior noise immunity compared to Rayleigh speckles and sub-Rayleigh speckles, as quantified by image PSNR and SSIM metrics.

\section{Conclusion}

In conclusion, we have investigated a scheme of GISC spectral camera with super-Rayleigh modulator. A flexible method for generating non-Rayleigh speckle patterns is proposed and verified by both simulation and experiment. By this method, non-Rayleigh speckle patterns can be realized in different speckle detection planes and central wavelengths according to the required system parameters. Simultaneously, simulation and experimental results indicate that the noise immunity of the system and the quality of the reconstructed results can be improved by using super-Rayleigh speckle patterns. In order to obtain super-Rayleigh speckles over a broader spectral range, equipping the GISC spectral camera with a flat-field grating~\cite{liu2018hyperspectral} may be a feasible method.
GISC spectral camera with super-Rayleigh modulator can be applied to many practical imaging areas, such as single photon imaging~\cite{liu2016single}, super-resolution imaging~\cite{chaigne20167} and GISC nanoscopy~\cite{li2019single}.

\appendix
\section*{Theoretical derivation}

Figure \ref{fig:speckle_generation} is the schematic of generating a phase matrix loaded onto SLM to obtain super-Rayleigh speckles. Suppose an exponential factor is added to Rayleigh speckle field $U_{\rm{Ray}}\left( {{{\bm{r_{1}}}}} \right)$ to obtain super-Rayleigh speckle field $W\left( {{\bm{r_{1}}}} \right) = {\left( {{U_{\rm{Ray}}}\left( {{\bm{r_{1}}}} \right)} \right)^n}$, where ${\bm{r_{1}}}$ represents a two-dimensional vector on the $xy$ plane. According to the principle of reversibility of light, the light field $W\left( {{\bm{r_{1}}}} \right)$ with wavelength $\lambda_1$ inversely propagates for a distance $z_2$ to the SLM plane, and the light field on the $\alpha\beta$ plane can be denoted as 
\begin{equation}\label{eq:a8}\tag{A1}
\begin{aligned}
U\left( {{{\bm{r_{0}}}},{\lambda _1}} \right) =& \frac{j}{{{\lambda _1}{z_2}}}\exp \left( { - j\frac{{2\pi }}{{{\lambda _1}}}{z_2}} \right)\int\limits_{ - \infty }^\infty  W\left( {{{\bm{r_{1}}}}} \right)\nonumber\\
&\times \exp \left( { - j\frac{\pi }{{{\lambda _1}{z_2}}}{{\left( {{{\bm{r_{0}}}} - {{\bm{r_{1}}}}} \right)}^2}} \right) d{{\bm{r_{1}}}},
\end{aligned}
\end{equation}
where $\bm{r_0}$ represents a two-dimensional vector on the $\alpha\beta$ plane. Then suppose that the transmittance function of the SLM is $T\left( {{{\bm{r_{0}}}},{\lambda _1}} \right) = U\left( {{{\bm{r_{0}}}},{\lambda _1}} \right)$.

To investigate the role of system parameters of GISC spectral camera in the formation of super-Rayleigh speckles, we denote a monochromatic point source at pixel $\bm{r_a}$ with wavelength $\lambda_2$ on the $x_0y_0$ plane (Fig. \ref{fig:system}) propagating to the SLM as ${U_0}\left( {\bm{r_0},{\lambda _2}} \right) \approx \exp \left( {{{j\pi {{\left( {\bm{r_0} - \bm{r_a}} \right)}^2}} \mathord{\left/
 {\vphantom {{j\pi {{\left( {\bm{r_0} - \bm{r_a}} \right)}^2}} {{\lambda _2}{z_1} + {{j2\pi {z_1}} \mathord{\left/
 {\vphantom {{j2\pi {z_1}} {{\lambda _2}}}} \right.
 \kern-\nulldelimiterspace} {{\lambda _2}}}}}} \right.
 \kern-\nulldelimiterspace} {{\lambda _2}{z_1} + {{j2\pi {z_1}} \mathord{\left/
 {\vphantom {{j2\pi {z_1}} {{\lambda _2}}}} \right.
 \kern-\nulldelimiterspace} {{\lambda _2}}}}}} \right)$ (based on the paraxial approximation). Then, under Fresnel diffraction theorem, the light field on the speckle plane locating at ${z_2}'$ behind the SLM can be calculated as 
\begin{equation}\label{eq:a9}\tag{A2}
\begin{aligned}
&U\left( {{{\bm{r_1}}},{\lambda _2}} \right)
= \frac{1}{{j{\lambda _2}{z_2}^\prime }}\exp \left( {j\frac{{2\pi }}{{{\lambda _2}}}{z_2}^\prime } \right)\int U\left( {{{\bm{r_0}}},{\lambda _2}} \right)T\left( {{{\bm{r_0}}},{\lambda _1}} \right)\\
&\times\exp\left( {j\frac{\pi }{{{\lambda _2}{z_2}^\prime }}{{\left( {{{\bm{r_1}}} - {{\bm{r_0}}}} \right)}^2}} \right){{{\rm{d}}\bm{r_0}}} \\
 &= \frac{1}{{{\lambda _2}{z_2}^\prime }{{\lambda _1}{z_2}}}\exp \left( j2\pi\left({ \frac{{z_2}^\prime }{{{\lambda _2}}} -  \frac{{z_2}}{{{\lambda _1}}} + \frac{{z_1}}{{{\lambda _2}}}}\right) \right)\exp \left( {\frac{j\pi{{\bm{r_1}}}^2 }{{{\lambda _2}{z_2}^\prime }}} \right)\\
& \times \int {W\left( {{{\bm{\xi }}}} \right)\exp \left( { \frac{- j\pi{\bm{\xi }}^2 }{{{\lambda _1}{z_2}}}} \right)\left[ {\int {\exp \left( { \frac{j2\pi{\bm{\xi }}{{\bm{r}}_{\bm{0}}}}{{{\lambda _1}{z_2}}}}- { \frac{- j2\pi {{{\bm{r_1}}}}{{\bm{r}}_{\bm{0}}}}{{{\lambda _2}{z_2}^\prime }}}\right)} } \right.}\\
&\left. { \times \exp \left( {j\frac{{\pi {{\bm{r}}_{\bm{0}}}^2}}{{{z_2}^\prime {\lambda _2}}} - j\frac{{\pi {{\bm{r}}_{\bm{0}}}^2}}{{{z_2}{\lambda _1}}} + j\frac{{\pi {{\left( {{{\bm{r}}_{\bm{0}}} - {{\bm{r}}_{\bm{a}}}} \right)}^2}}}{{{\lambda _2}{z_1}}}} \right){\rm{d}}{{\bm{r}}_{\bm{0}}}} \right]{\rm{d}}\bm{\xi }\\
&= j\chi\exp \left( j2\pi\left({ \frac{{z_2}\beta }{{\gamma {\lambda _1}}} - \frac{{z_2}}{{{\lambda _1}}} + \frac{{z_1}\beta }{{{\lambda _1}}}} \right)\right)\exp \left( {j\frac{{\beta \gamma \pi }}{{{\lambda _1}{z_2}}}{{\bm{r}}_{\bm{1}}}^2} \right)\\
 &\times \exp \left( {j\frac{\pi }{{{\lambda _1}}}\left( {\frac{{\beta {{\bm{r}}_{\bm{a}}}^2}}{{{z_1}}} - \frac{{{\beta }{z_2}\chi{{\bm{r}}_{\bm{a}}}^2}}{{{z_{1}}^2}\gamma}} \right)} \right)\exp \left( {\frac{ - j2\pi {\beta{{\bm{r}}_{\bm{a}}}{\bm{r_1}}}}{{{\lambda _1}z_1}}} \right)\\
 &\times \int W\left( {\bm{\xi }} \right)\exp \left( { \frac{- j\pi }{{{\lambda _1}{z_2}}}{{\bm{\xi }}^2}} \right)\exp\left( {\frac{-j\pi\chi}{\gamma\lambda_1 z_2\beta}} \right.\\
&\times\left. {\left( {{\beta ^2}{\gamma ^2}{{\bm{r}}_{\bm{1}}}^2 - 2\beta \gamma {{\bm{r_1}}}{\bm{\xi }} + {{\bm{\xi }}^2}}-2\tau\beta\bm{r_a}\bm{\xi} \right)} \right)\rm{d}\bm{\xi},
\end{aligned}
\vspace{8pt}
\end{equation}
where $\chi {\rm{ = }}{{\beta \gamma {z_1}} \mathord{\left/
 {\vphantom {{\beta \gamma {z_1}} {\left( {\left( {\beta \gamma  - 1} \right){z_1} + \beta {z_2}} \right)}}} \right. \kern-\nulldelimiterspace} {\left( {\left( {\beta \gamma  - 1} \right){z_1} + \beta {z_2}} \right)}}$, $\beta  = {{{\lambda _1}} \mathord{\left/ {\vphantom {{{\lambda _1}} {{\lambda _2}}}} \right. \kern-\nulldelimiterspace} {{\lambda _2}}}$, $\gamma  = {{{{z_2}} \mathord{\left/ {\vphantom {{{z_2}} {{z_2}}}} \right. \kern-\nulldelimiterspace} {{z_2}}}^\prime }$, $\tau  = {{{z_2}} \mathord{\left/ {\vphantom {{{z_2}} {{z_1}}}} \right.\kern-\nulldelimiterspace} {{z_1}}}$. 
The contrast of speckles is defined as ${g^2} = {{\left\langle {I{{\left( {{{\bm{r}}_{\bm{1}}},{\lambda _2}} \right)}^2}} \right\rangle } \mathord{\left/
 {\vphantom {{\left\langle {I{{\left( {{{\bm{r}}_{\bm{1}}},{\lambda _2}} \right)}^2}} \right\rangle } {{{\left\langle {I\left( {{{\bf{r}}_{\bf{1}}},{\lambda _2}} \right)} \right\rangle }^2}}}} \right.
 \kern-\nulldelimiterspace} {{{\left\langle {I\left( {{{\bm{r}}_{\bm{1}}},{\lambda _2}} \right)} \right\rangle }^2}}} - 1$, where $\left\langle...\right\rangle$ denotes ensemble averaging. Hence, according to Eq. (\ref{eq:a9}), we have 
\begin{equation}\label{eq:a13}\tag{A3}
\begin{aligned}
&\left\langle {{I^2}\left( {{{\bm{r_1}}},{\lambda _2}} \right)} \right\rangle 
= \left\langle {{{\left( {U\left( {{{\bm{r_1}}},{\lambda _2}} \right){U^*}\left( {{{\bm{r_1}}},{\lambda _2}} \right)} \right)}^2}} \right\rangle\\
&={\chi^4}\iiiint {{{{\left\langle {W\left( {\bm{\xi }} \right){W^*}\left( {{\bm{\xi '}}} \right)W\left( {{\bm{\xi ''}}} \right){W^*}\left( {{\bm{\xi '''}}} \right)} \right\rangle}}}}\\ &\times\exp\left( {\frac{- j\pi \chi}{\gamma\lambda_1 z_2}} \right. \left( {\left(\gamma+\tau \right)\left( {{{\bm{\xi }}^2} - {{{\bm{\xi '}}}^2} + {{{\bm{\xi ''}}}^2} - {{{\bm{\xi '''}}}^2}} \right)} \right.\\
&\left. {\left. { - 2\left( {{\bm{r_1}}}\gamma+\tau{\bm{r_a}} \right)\left( {{\bm{\xi }} - {\bm{\xi '}} + {\bm{\xi ''}} - {\bm{\xi '''}}} \right)} \right)} \right){\rm{d}}{\bm{\xi }}{\rm{d}}{\bm{\xi '}}{\rm{d}}{\bm{\xi ''}}{\rm{d}}{\bm{\xi '''}}
\end{aligned}
\vspace{8pt}
\end{equation}
and
\begin{equation}\tag{A4}
\begin{aligned}\label{eq:a14}
&{\left\langle {I\left( {{{\bm{r_1}}},{\lambda _2}} \right)} \right\rangle ^2}
={\left\langle {U\left( {{{\bm{r_1}}},{\lambda _2}} \right){U^*}\left( {{{\bm{r_1}}},{\lambda _2}} \right)} \right\rangle ^2}\\
&={\chi^4}\iiiint {{{{\left\langle {W\left( {\bm{\xi }} \right){W^*}\left( {{\bm{\xi '}}} \right)} \right\rangle \left\langle {W\left( {{\bm{\xi ''}}} \right){W^*}\left( {{\bm{\xi '''}}} \right)} \right\rangle }}}}\\
&\times\exp\left( {\frac{- j\pi \chi}{\gamma\lambda_1 z_2}} \right. \left( {\left(\gamma+\tau \right)\left( {{{\bm{\xi }}^2} - {{{\bm{\xi '}}}^2} + {{{\bm{\xi ''}}}^2} - {{{\bm{\xi '''}}}^2}} \right)} \right.\\
&\left. {\left. { - 2\left( {{\bm{r_1}}}\gamma+\tau{\bm{r_a}} \right)\left( {{\bm{\xi }} - {\bm{\xi '}} + {\bm{\xi ''}} - {\bm{\xi '''}}} \right)} \right)} \right){\rm{d}}{\bm{\xi }}{\rm{d}}{\bm{\xi '}}{\rm{d}}{\bm{\xi ''}}{\rm{d}}{\bm{\xi '''}}.
\end{aligned}
\vspace{8pt}
\end{equation} 
Here for convenience, $n \in {N^*}$ is considered. Suppose that Rayleigh speckle field is satisfied with a complex-valued circular Gaussian random process. Then according to the Gauss moments theorem, the ensemble averaging about light field $W\left(\bm{\xi}\right)$ in Eqs. (\ref{eq:a13}) and (\ref{eq:a14}), ${\left\langle {W\left( {\bm{\xi }} \right){W^*}\left( {{\bm{\xi '}}} \right)W\left( {{\bm{\xi ''}}} \right){W^*}\left( {{\bm{\xi '''}}} \right)} \right\rangle }$ and ${\left\langle {W\left( {\bm{\xi }} \right){W^*}\left( {{\bm{\xi '}}} \right)} \right\rangle \left\langle {W\left( {{\bm{\xi ''}}} \right){W^*}\left( {{\bm{\xi '''}}} \right)} \right\rangle }$, can be represented as 
\begin{equation}\tag{A5}
\begin{aligned}\label{eq:a15}
&\left\langle {W\left( {\bm{\xi }} \right){W^*}\left( {{\bm{\xi '}}} \right)W\left( {{\bm{\xi ''}}} \right){W^*}\left( {{\bm{\xi '''}}} \right)} \right\rangle \nonumber\\
&=\left\langle {{{\left( {{U_{\rm{Ray}}}\left( {\bm{\xi }} \right)} \right)}^n}{{\left( {U_{\rm{Ray}}^*\left( {{\bm{\xi '}}} \right)} \right)}^n}{{\left( {{U_{\rm{Ray}}}\left( {{\bm{\xi ''}}} \right)} \right)}^n}{{\left( {U_{\rm{Ray}}^*\left( {{\bm{\xi '''}}} \right)} \right)}^n}} \right\rangle \nonumber\\
& = {\left( {n!} \right)^2}{\left\langle {{U_{\rm{Ray}}}\left( {\bm{\xi }} \right)U_{\rm{Ray}}^*\left( {{\bm{\xi '}}} \right)} \right\rangle ^n}{\left\langle {{U_{\rm{Ray}}}\left( {{\bm{\xi ''}}} \right)U_{\rm{Ray}}^*\left( {{\bm{\xi '''}}} \right)} \right\rangle ^n}\nonumber\\
&+ {\left( {n!} \right)^2}{\left\langle {{U_{\rm{Ray}}}\left( {\bm{\xi }} \right)U_{\rm{Ray}}^*\left( {{\bm{\xi '''}}} \right)} \right\rangle ^n}{\left\langle {{U_{\rm{Ray}}}\left( {{\bm{\xi ''}}} \right)U_{\rm{Ray}}^*\left( {{\bm{\xi '}}} \right)} \right\rangle ^n}\nonumber\\
&+ \sum\limits_{k = 1}^{n - 1}\left( {{{{\left( {C_n^k} \right)}^4}{{\left( {k!\left( {n - k} \right)!} \right)}^2}}} \right.\nonumber\\
&\times {\left\langle {{U_{\rm{Ray}}}\left( {\bm{\xi }} \right)U_{\rm{Ray}}^*\left( {{\bm{\xi '}}} \right)} \right\rangle ^k}{\left\langle {{U_{\rm{Ray}}}\left( {\bm{\xi }} \right)U_{\rm{Ray}}^*\left( {{\bm{\xi '''}}} \right)} \right\rangle ^{n - k}}\nonumber\\
&\left. {\times {\left\langle {{U_{\rm{Ray}}}\left( {{\bm{\xi ''}}} \right)U_{\rm{Ray}}^*\left( {{\bm{\xi '}}} \right)} \right\rangle ^{n - k}}{\left\langle {{U_{\rm{Ray}}}\left( {{\bm{\xi ''}}} \right)U_{\rm{Ray}}^*\left( {{\bm{\xi '''}}} \right)} \right\rangle ^k}} \right)
\end{aligned}
\vspace{8pt}
\end{equation}
and
\begin{equation}\label{eq:a16}\tag{A6}
\begin{aligned}
&\left\langle {W\left( {\bm{\xi }} \right){W^*}\left( {{\bm{\xi '}}} \right)} \right\rangle \left\langle {W\left( {\bm{\xi }} \right){W^*}\left( {{\bm{\xi '}}} \right)} \right\rangle\\ &=\left\langle {{{\left( {{U_{\rm{Ray}}}\left( {\bm{\xi }} \right)} \right)}^n}{{\left( {U_{\rm{Ray}}^*\left( {{\bm{\xi '}}} \right)} \right)}^n}} \right\rangle \left\langle {{{\left( {{U_{\rm{Ray}}}\left( {{\bm{\xi ''}}} \right)} \right)}^n}{{\left( {U_{\rm{Ray}}^*\left( {{\bm{\xi '''}}} \right)} \right)}^n}} \right\rangle \\
& = {\left( {n!} \right)^2}{\left\langle {{U_{\rm{Ray}}}\left( {\bm{\xi }} \right)U_{\rm{Ray}}^*\left( {{\bm{\xi '}}} \right)} \right\rangle ^n}{\left\langle {{U_{\rm{Ray}}}\left( {{\bm{\xi ''}}} \right)U_{\rm{Ray}}^*\left( {{\bm{\xi '''}}} \right)} \right\rangle ^n}.
\end{aligned}
\vspace{8pt}
\end{equation} 
Furthermore, assuming that the autocorrelation of Rayleigh speckle field ${U_{\rm{Ray}}}\left( {{{\bm{r_{1}}}}} \right)$ obeys Gaussian distribution which can be expressed as $ \left\langle {{U_{\rm{Ray}}}\left( {\bm{r_{1}}} \right)U_{\rm{Ray}}^*\left( {{\bm{r_{1} '}}} \right)} \right\rangle  =  {{\exp \left\{ { - {{{{\left( {{\bm{r_{1} }} - {\bm{r_{1} '}}} \right)}^2}} \mathord{\left/
 {\vphantom {{{{\left( {{\bm{r_{1} }} - {\bm{r_{1} '}}} \right)}^2}} {2{\sigma ^2}}}} \right.
 \kern-\nulldelimiterspace} {2{\sigma ^2}}}} \right\}} \mathord{\left/
 {\vphantom {{\exp \left\{ { - {{{{\left( {{\bm{r_{1} }} - {\bm{r_{1} '}}} \right)}^2}} \mathord{\left/
 {\vphantom {{{{\left( {{\bm{r_{1} }} - {\bm{r_{1} '}}} \right)}^2}} {2{\sigma ^2}}}} \right.
 \kern-\nulldelimiterspace} {2{\sigma ^2}}}} \right\}} {\sqrt {2\pi \sigma } }}} \right.
 \kern-\nulldelimiterspace} {\sqrt {2\pi \sigma } }}$ where $\sigma$ is the standard deviation, and substituting Eqs. (\ref{eq:a13})-(\ref{eq:a16}) into the expression of contrast, we have 
\begin{equation}\label{eq:a17}\tag{A7}
\begin{aligned}
{g^2} =& 1 + \frac{{\sum\limits_{k = 1}^{n - 1} {{{\left( {C_n^k} \right)}^4}{{\left( {k!\left( {n - k} \right)!} \right)}^2}} }}{{{{\left( {n!} \right)}^2}}}\\
&\times \frac{\begin{array}{c}
\iiiint{{{{{{\left\langle {{U_{{\rm{Ray}}}}\left( {\bm{\xi }} \right)U_{{\rm{Ray}}}^*\left( {{\bm{\xi '}}} \right)} \right\rangle }^k}} } } }\vspace{5pt} \\
 \qquad\qquad\times {\left\langle {{U_{{\rm{Ray}}}}\left( {\bm{\xi }} \right)U_{{\rm{Ray}}}^*\left( {{\bm{\xi '''}}} \right)} \right\rangle ^{n - k}}\vspace{5pt}\\
 \qquad\qquad\times {\left\langle {{U_{{\rm{Ray}}}}\left( {{\bm{\xi ''}}} \right)U_{{\rm{Ray}}}^*\left( {{\bm{\xi '}}} \right)} \right\rangle ^{n - k}}\vspace{5pt}\\
 \qquad\qquad\times {\left\langle {{U_{{\rm{Ray}}}}\left( {{\bm{\xi ''}}} \right)U_{{\rm{Ray}}}^*\left( {{\bm{\xi '''}}} \right)} \right\rangle ^k}\vspace{5pt}\\
 \qquad\qquad\times \exp \left(  \cdots  \right){\rm{d}}{\bm{\xi }}{\rm{d}}{\bm{\xi '}}{\rm{d}}{\bm{\xi ''}}{\rm{d}}{\bm{\xi '''}}
\end{array}}{\begin{array}{l}
\iiiint {{{{{{\left\langle {{U_{{\rm{Ray}}}}\left( {\bm{\xi }} \right)U_{{\rm{Ray}}}^*\left( {{\bm{\xi '}}} \right)} \right\rangle }^n}} } } }\vspace{5pt} \\
 \qquad\times {\left\langle {{U_{{\rm{Ray}}}}\left( {{\bm{\xi ''}}} \right)U_{{\rm{Ray}}}^*\left( {{\bm{\xi '''}}} \right)} \right\rangle ^n}\vspace{5pt}\\
 \qquad\times \exp \left(  \cdots  \right){\rm{d}}{\bm{\xi }}{\rm{d}}{\bm{\xi '}}{\rm{d}}{\bm{\xi ''}}{\rm{d}}{\bm{\xi '''}}
\end{array}}\\
 =& 1 + \frac{{\sum\limits_{k = 1}^{n - 1} {{{\left( {C_n^k} \right)}^4}{{\left( {k!\left( {n - k} \right)!} \right)}^2}} }}{{{{\left( {n!} \right)}^2}}}\\
& \times \frac{\begin{array}{l}
\iiiint {{{ {\exp ( - \frac{{k{{\left( {{\bm{\xi }} - {\bm{\xi '}}} \right)}^2}}}{{2{\sigma ^2}}})\exp ( - \frac{{\left( {n - k} \right){{\left( {{\bm{\xi }} - {\bm{\xi '''}}} \right)}^2}}}{{2{\sigma ^2}}})} } } } \vspace{5pt}\\
 \qquad\times \exp ( - \frac{{\left( {n - k} \right){{\left( {{\bm{\xi ''}} - {\bm{\xi '}}} \right)}^2}}}{{2{\sigma ^2}}})\exp ( - \frac{{k{{\left( {{\bm{\xi ''}} - {\bm{\xi '''}}} \right)}^2}}}{{2{\sigma ^2}}})\vspace{5pt}\\
 \qquad\times \exp \left(  \cdots  \right){\rm{d}}{\bm{\xi }}{\rm{d}}{\bm{\xi '}}{\rm{d}}{\bm{\xi ''}}{\rm{d}}{\bm{\xi '''}}
\end{array}}{\begin{array}{l}
\iiiint {{{{\exp \left( { - \frac{{n{{\left( {{\bm{\xi }} - {\bm{\xi '}}} \right)}^2}}}{{{\sigma ^2}}}} \right)\exp \left( { - \frac{{n{{\left( {{\bm{\xi ''}} - {\bm{\xi '''}}} \right)}^2}}}{{{\sigma ^2}}}} \right)} } } } \vspace{5pt}\\
 \qquad\times \exp \left(  \cdots  \right){\rm{d}}{\bm{\xi }}{\rm{d}}{\bm{\xi '}}{\rm{d}}{\bm{\xi ''}}{\rm{d}}{\bm{\xi '''}}
\end{array}}
\end{aligned}
\vspace{8pt}
\end{equation}
with 
\begin{equation}\label{eq:a18}\tag{A8}
\begin{aligned}
 &\exp \left(  \cdots  \right)\\
&= \exp\left( {\frac{- j\pi\chi }{\gamma{\lambda _1}{z_2}}} \right. \left( {\left(\gamma+\tau\right)\left({{{\bm{\xi }}^2} - {{{\bm{\xi '}}}^2} + {{{\bm{\xi ''}}}^2} - {{{\bm{\xi '''}}}^2}}\right)} \right.\\
&\quad\left. {\left. { -2\left( {{\bm{r_1}}}\gamma+\tau{\bm{r_a}}\right)} \left({{\bm{\xi }} - {\bm{\xi '}} + {\bm{\xi ''}} - {\bm{\xi '''}}} \right)\right)} \right). 
 \end{aligned}
 \vspace{8pt}
\end{equation}
Next, we can get the final expression of the contrast by calculating the following two integrals
\begin{equation}\label{eq:a19}\tag{A9}
\begin{aligned}
&\iiiint {{{{\exp \left( { - \frac{{n{{\left( {{\bm{\xi }} - {\bm{\xi '}}} \right)}^2}}}{{{\sigma ^2}}}} \right)\exp \left( { - \frac{{n{{\left( {{\bm{\xi ''}} - {\bm{\xi '''}}} \right)}^2}}}{{{\sigma ^2}}}} \right)} } } }\\
&\qquad\quad\times\exp\left(\cdots\right){\rm{d}}{\bm{\xi }}{\rm{d}}{\bm{\xi '}}{\rm{d}}{\bm{\xi ''}}{\rm{d}}{\bm{\xi '''}} \vspace{5pt}\\
&\mathop  = \limits^{{\bm{\xi }} - {\bm{\xi '}} = {\Delta\bm{ \alpha }}}\left( {\iint \exp\left(-\frac{n\Delta\bm{\alpha}^2}{\sigma^2}\right)} \right.\exp \left( {\frac{-j\pi \chi}{\gamma \lambda_1 z_2}} \right.\\
&\quad\times\left( {\left( \gamma +\tau\right)\left(2\bm{\xi}\Delta\bm{\alpha}-{\Delta \alpha}^2 \right)} \right.\\
&\quad\left. {\left. {\left. {-2\left(\bm{r_1}\gamma+\tau\bm{r_a} \right)\Delta\bm{\alpha}} \right)} \right)} {\rm{d}}{\bm{\xi }}{\rm{d}}{{\Delta\bm{ \alpha }}} \right)^2\\
&= {\left( {\frac{{{\lambda _1}\gamma {z_1}{z_2}}}{{\chi \left( {\gamma {z_1} + {z_2}} \right)}}} \right)^4}
\end{aligned} 
\end{equation}
and
\begin{equation}\label{eq:a20}\tag{A10}
\begin{aligned}
&\iiiint {{{{\exp \left( { - \frac{{k{{\left( {{\bm{\xi }} - {\bm{\xi '}}} \right)}^2}}}{{2{\sigma ^2}}}} \right)\exp \left( { - \frac{{\left( {n - k} \right){{\left( {{\bm{\xi }} - {\bm{\xi '''}}} \right)}^2}}}{{2{\sigma ^2}}}} \right)} } } } \\
 &\times \exp \left( { - \frac{{\left( {n - k} \right){{\left( {{\bm{\xi ''}} - {\bm{\xi '}}} \right)}^2}}}{{2{\sigma ^2}}}} \right)\exp \left( { - \frac{{k{{\left( {{\bm{\xi ''}} - {\bm{\xi '''}}} \right)}^2}}}{{2{\sigma ^2}}}} \right)\\
 &\times\exp \left(  \cdots  \right){\rm{d}}{\bm{\xi }}{\rm{d}}{\bm{\xi '}}{\rm{d}}{\bm{\xi ''}}{\rm{d}}{\bm{\xi '''}}\\
 &= \iiiint {{ {{\exp \left( - \frac{{k{{\left( {{\bm{\xi }} - {\bm{\xi '}}} \right)}^2}}}{{2{\sigma ^2}}}\right)\exp \left( - \frac{{\left( {n - k} \right){{\left( {{\bm{\xi }} - {\bm{\xi '''}}} \right)}^2}}}{{2{\sigma ^2}}}\right)} } } }\\
 &\times\exp \left( - \frac{{\left( {n - k} \right){{\left( {{\bm{\xi ''}} - {\bm{\xi '}}} \right)}^2}}}{{2{\sigma ^2}}}\right)\exp \left( - \frac{{k{{\left( {{\bm{\xi ''}} - {\bm{\xi '''}}} \right)}^2}}}{{2{\sigma ^2}}}\right)\\
 &\times \exp \left( {\frac{-j\pi\chi}{\gamma \lambda_1 z_2}} \right.\left( {{\frac{\gamma z_1+z_2}{z_1}}} \right.\left( {\left( {{\bm{\xi}-\frac{{{z_1}\left(\gamma {{\bm{r_1}}}+\tau{{\bm{r_a}}}\right)}}{{\gamma {z_1} + {z_2}}}}} \right)^2} \right.\\
&-\left( {{\bm{\xi'}-\frac{{{z_1}\left(\gamma {{\bm{r_1}}}+\tau{{\bm{r_a}}}\right)}}{{\gamma {z_1} + {z_2}}}}} \right)^2+\left( {\bm{\xi''}-\frac{{{z_1}\left(\gamma {{\bm{r_1}}}+\tau{{\bm{r_a}}}\right)}}{{\gamma {z_1} + {z_2}}}}\right)^2\\
&\left. {\left. {\left. {-\left( {{\bm{\xi'''}-\frac{{{z_1}\left(\gamma {{\bm{r_1}}}+\tau{{\bm{r_a}}}\right)}}{{\gamma {z_1} + {z_2}}}}} \right)^2} \right)} \right)} \right){\rm{d}}{\bm{\xi }}{\rm{d}}{\bm{\xi '}}{\rm{d}}{\bm{\xi ''}}{\rm{d}}{\bm{\xi '''}}.
\end{aligned}
\vspace{8pt}
\end{equation}
Then we utilize the following variable substitution
\begin{subequations}
\label{eq:whole1}
\begin{equation}\label{eq:a21}\tag{A11a}
{\bm{\xi }} - \frac{{{z_1}\gamma \left({{\bm{r_1}}}+\tau{{\bm{r_a}}}\right)}}{{\gamma {z_1} + {z_2}}} = {\bm{\zeta }},
\end{equation}
\begin{equation}\label{eq:a22}\tag{A11b}
{\bm{\xi '}} - \frac{{{z_1}\gamma \left({{\bm{r_1}}}+\tau{{\bm{r_a}}}\right)}}{{\gamma {z_1} + {z_2}}} = {\bm{\zeta '}},
\end{equation}
\begin{equation}\label{eq:a23}\tag{A11c}
{\bm{\xi ''}} - \frac{{{z_1}\gamma \left({{\bm{r_1}}}+\tau{{\bm{r_a}}}\right)}}{{\gamma {z_1} + {z_2}}} = {\bm{\zeta ''}},
\end{equation}
\begin{equation}\label{eq:a24}\tag{A11d}
{\bm{\xi '''}} - \frac{{{z_1}\gamma \left({{\bm{r_1}}}+\tau{{\bm{r_a}}}\right)}}{{\gamma {z_1} + {z_2}}} = {\bm{\zeta '''}},
\end{equation}
\end{subequations}
and 
\begin{subequations}
\label{eq:whole2}
\begin{equation}\label{eq:a25}\tag{A12a}
{\bm{\zeta }} - {\bm{\zeta '}} = {\Delta\bm{ \zeta }},
\end{equation}
\begin{equation}\label{eq:a26}\tag{A12b}
 {\bm{\zeta }} - {\bm{\zeta '''}} = {\Delta\bm{ \zeta '}},
\end{equation}
\begin{equation}\label{eq:a27}\tag{A12c}
{\bm{\zeta ''}} - {\bm{\zeta '}} = {\Delta\bm{ \zeta ''}},
\end{equation}
\begin{equation}\label{eq:a28}\tag{A12d}
{\bm{\zeta ''}} - {\bm{\zeta '''}} = {\Delta {\bm{\zeta '}} - \Delta {\bm{\zeta }} + \Delta {\bm{\zeta ''}}}.
\end{equation}
\end{subequations}

Taking Eqs. (\ref{eq:a21})-(\ref{eq:a28}) into Eq. (\ref{eq:a20}) yields
\begin{equation}\label{eq:a29}\tag{A13}
\begin{aligned}
&\iiiint {{{{\exp \left( { - \frac{{k{{\left( {{\bm{\xi }} - {\bm{\xi '}}} \right)}^2}}}{{2{\sigma ^2}}}} \right)\exp \left( { - \frac{{\left( {n - k} \right){{\left( {{\bm{\xi }} - {\bm{\xi '''}}} \right)}^2}}}{{2{\sigma ^2}}}} \right)} } } }\nonumber \\
&\times \exp \left( { - \frac{{\left( {n - k} \right){{\left( {{\bm{\xi ''}} - {\bm{\xi '}}} \right)}^2}}}{{2{\sigma ^2}}}} \right)\exp \left( { - \frac{{k{{\left( {{\bm{\xi ''}} - {\bm{\xi '''}}} \right)}^2}}}{{2{\sigma ^2}}}} \right)\nonumber\\
&\times\exp \left(  \cdots  \right){\rm{d}}{\bm{\xi }}{\rm{d}}{\bm{\xi '}}{\rm{d}}{\bm{\xi ''}}{\rm{d}}{\bm{\xi '''}}\nonumber\\
 &={\left( \frac{\lambda_1 \gamma z_1 z_2}{\chi \left(\gamma z_1 +z_2\right)} \right)^2}\nonumber\\
 &\times\frac{{{\pi ^2}{\sigma ^4}{\lambda _1}^2{z_2}^2{z_1}^2\beta^2\gamma^2}}{{k\left( {n - k} \right){\lambda _1}^2{z_2}^2{z_1}^2\beta^2\gamma^2 + \chi^2{\pi ^2}{\sigma ^4}{\beta ^2}{{\left( {\gamma {z_1} + {z_2}} \right)}^2}}}. 
\end{aligned}
\vspace{8pt}
\end{equation}
Substituting Eqs. (\ref{eq:a19}) and (\ref{eq:a29}) into Eq. (\ref{eq:a17}), the contrast $g^2$ of super-Rayleigh speckle field can be ultimately expressed as 
\begin{equation}\label{eq:a30}\tag{A14}
{g^2} = 1 + \sum\limits_{k = 1}^{n - 1} \frac{{{{ {{{\left( {C_n^k} \right)}^4}{{\left( {k!\left( {n - k} \right)!} \right)}^2}} } \mathord{\left/
 {\vphantom {{\sum\limits_{k = 1}^{n - 1} {{{\left( {C_n^k} \right)}^4}{{\left( {k!\left( {n - k} \right)!} \right)}^2}} } {{{\left( {n!} \right)}^2}}}} \right.
 \kern-\nulldelimiterspace} {{{\left( {n!} \right)}^2}}}}}{{1 + {{k\left( {n - k} \right){z_2}^2{{\left( {{\lambda _1} - \frac{1}{{\left( {\gamma  + \tau } \right)}}{\lambda _2}} \right)}^2}} \mathord{\left/
 {\vphantom {{k\left( {n - k} \right){z_2}^2{{\left( {{\lambda _1} - \frac{1}{{\left( {\gamma  + \tau } \right)}}{\lambda _2}} \right)}^2}} {{\pi ^2}{\sigma ^4}}}} \right.
 \kern-\nulldelimiterspace} {{\pi ^2}{\sigma ^4}}}}}.
 \vspace{8pt}
\end{equation}
From Eq. (\ref{eq:a30}), we can see that the contrast of super-Rayleigh speckles is determined by the parameters $z_1$, $z_2$, ${z_2}'$, $\lambda_1$, $\lambda_2$, and $n$.


%

\ifCLASSOPTIONcaptionsoff
  \newpage
\fi

\bibliographystyle{IEEEtran}
\bibliography{IEEEabrv,mybibfile}

\begin{IEEEbiography}[{\includegraphics[width=1in,height=1.25in,clip,keepaspectratio]{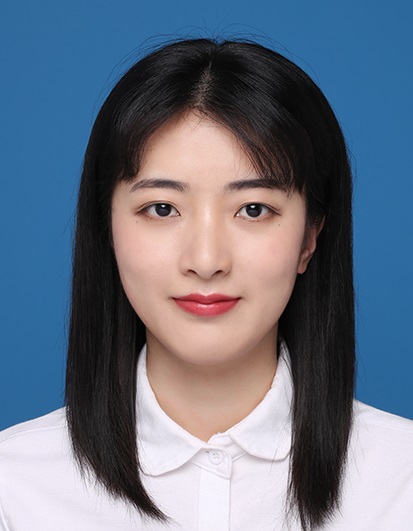}}]{Shengying Liu}
received the B.S. degree from the College of Mathematical Engineering from Zhejiang Normal University, Zhejiang, China, in 2014, and the M.S. degree from the Quantum Optics Laboratory, Shanghai Institute of Optics and Fine Mechanics, Shanghai, China, where she is currently pursuing the Ph.D. degree.

Her research interests include computational imaging, ghost imaging and snapshot spectral imaging.
\end{IEEEbiography}

\begin{IEEEbiography}[{\includegraphics[width=1in,height=1.25in,clip,keepaspectratio]{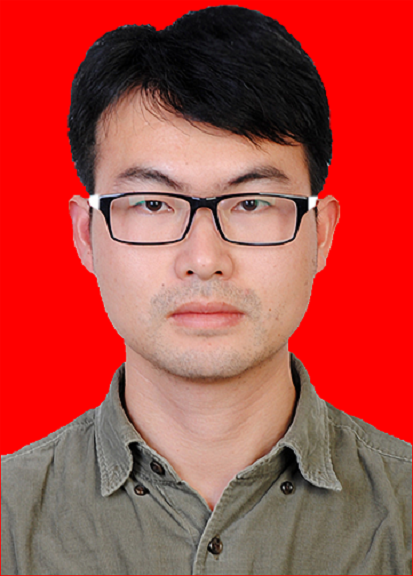}}]{Zhentao Liu}
received the B.S. degree in science and engineering from Nankai University and Tianjin University, Tianjin, China, in 2012, and the Ph.D. degree in optics from the Shanghai Institute of Optics and Fine Mechanics, CAS, Shanghai, China, in 2017.

He is currently an Assistant Researcher with the Shanghai Institute of Optics and Fine Mechanics,
CAS. His research interests include ghost imaging, high-order correlation imaging, imaging through
scattering media, super-resolution imaging, and imaging optimization.
\end{IEEEbiography}

\begin{IEEEbiography}[{\includegraphics[width=1in,height=1.25in,clip,keepaspectratio]{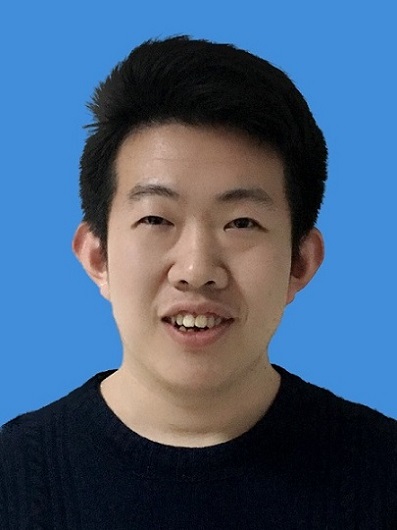}}]{Chenyu Hu}
received the B.S. degree in optical information science and technology from the University of Science and Technology of China, Hefei, China, in 2015. He is currently pursuing the Ph.D. degree with Shanghai Institute of Optics and Fine Mechanics, Chinese Academy of Sciences and University of Chinese Academy of Sciences, Shanghai, China.

His research interests include ghost imaging, computational imaging, and image processing.
\end{IEEEbiography}

\begin{IEEEbiography}[{\includegraphics[width=1in,height=1.25in,clip,keepaspectratio]{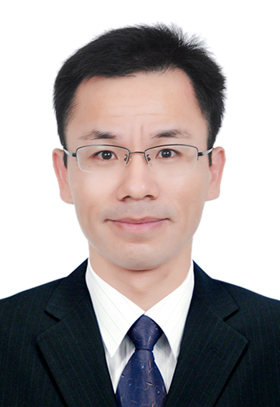}}]{Enrong Li}
received the B.S. degree in applied optics from Nankai University, Tianjin, China, in 2004, and the Ph.D. degree in optics from the Institute of High Energy Physics, Chinese Academy of Sciences, Beijing, China, in 2009.

He is currently an Associate Professor with the Shanghai Institute of Optics and Fine Mechanics,
Chinese Academy of Sciences, Shanghai, China. His research interests include image restoration,
compressive sensing algorithms for image reconstruction, and modeling of complex optical systems.
\end{IEEEbiography}

\begin{IEEEbiography}[{\includegraphics[width=1in,height=1.25in,clip,keepaspectratio]{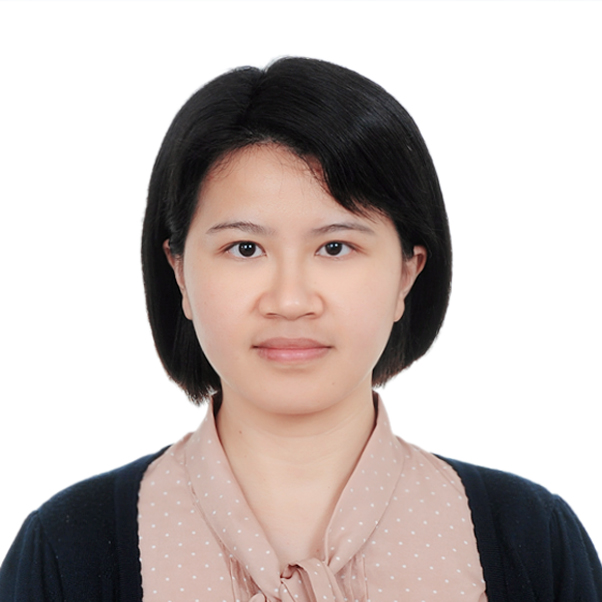}}]{Xia Shen}
received the B.S. degree in applied optics from Guangxi University, Nanning, China, in 2003, and the Ph.D. degree in optics from the Shanghai Institute of Optics and Fine Mechanics, Chinese Academy of Sciences, Shanghai, China, in 2008.

She is currently an Associate Professor with the Shanghai Institute of Optics and Fine Mechanics,
Chinese Academy of Sciences. Her research interests include computational imaging, ghost imaging, and spectral imaging.
\end{IEEEbiography}

\begin{IEEEbiography}[{\includegraphics[width=1in,height=1.25in,clip,keepaspectratio]{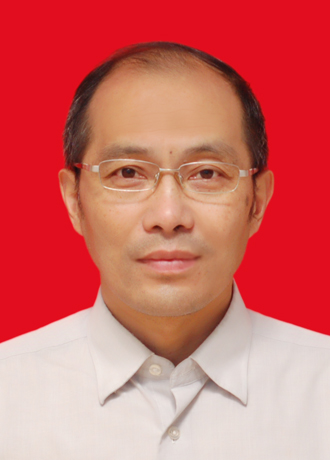}}]{Shensheng Han}
received the B.S. degree in physics and the Ph.D. degree in plasma physics from the University of Science and Technology of China, Hefei, China, in 1983 and 1990, respectively.

He is currently a Professor with the Shanghai Institute of Optics and Fine Mechanics, Chinese
Academy of Sciences. His research interests include plasma physics and computational optical
imaging based on high-order correlation of light fields. He is currently the Council Member of the Chinese Society for Optical Engineering and a member of Committee on the Quantum Optics of Chinese
Physical Society.
\end{IEEEbiography}







\end{document}